\documentclass[%
 reprint,
 amsmath,amssymb,
aip,
,onecolumn]{revtex4-2}

\usepackage[utf8]{inputenc}
\usepackage[T1]{fontenc}
\usepackage{hyperref}
\usepackage{url}
\usepackage{booktabs}
\usepackage{amsfonts}
\usepackage{nicefrac}
\usepackage{microtype}
\usepackage{changepage}
\usepackage{amsmath,amsfonts,amsthm,bm}
\usepackage{pgfplots}
\usepackage{float}
\usepackage{tikz}
\usepackage{caption}
\usepackage{subfig}
\usepackage{setspace}
\usepackage{amssymb}
\usepackage{dcolumn}
\usepackage{siunitx}
\usepackage{mathtools}
\usepackage{multirow}
\usepackage{parskip}
\usepgfplotslibrary{groupplots}
\usetikzlibrary{calc,decorations.markings}
\PassOptionsToPackage{dvipsnames,svgnames}{xcolor}
\usetikzlibrary{arrows,calc}
\usetikzlibrary{shapes.geometric}
\makeatletter%
\usepackage[toc,page]{appendix}
\usetikzlibrary{3d,calc}
\usepackage[section]{placeins}
\usetikzlibrary{external}
\makeatletter
\newcommand{\vast}{\bBigg@{4}}
\newcommand{\Vast}{\bBigg@{5}}
\makeatother

\tikzstyle{startstop} = [rectangle, rounded corners, minimum width=3cm, minimum height=1cm,text centered, draw=black, fill=red!0]
\tikzstyle{io} = [trapezium, trapezium left angle=70, trapezium right angle=110, minimum width=3cm, minimum height=1cm, text centered, text width=2cm, draw=black, fill=blue!0]
\tikzstyle{bigio} = [trapezium, trapezium left angle=70, trapezium right angle=110, minimum width=3cm, minimum height=1cm, text centered, text width=8cm, draw=black, fill=blue!0]
\tikzstyle{process} = [rectangle, minimum width=3cm, minimum height=1cm, text width=3cm, text centered, draw=black, fill=orange!0]
\tikzstyle{decision} = [diamond, minimum width=2cm, minimum height=0.5cm, aspect=2, text width=2cm, text centered, draw=black, fill=green!0]
\tikzstyle{arrow} = [thick,->,>=stealth]

\tikzset{
  on each segment/.style={
    decorate,
    decoration={
      show path construction,
      moveto code={},
      lineto code={
        \path [#1]
        (\tikzinputsegmentfirst) -- (\tikzinputsegmentlast);
      },
      curveto code={
        \path [#1] (\tikzinputsegmentfirst)
        .. controls
        (\tikzinputsegmentsupporta) and (\tikzinputsegmentsupportb)
        ..
        (\tikzinputsegmentlast);
      },
      closepath code={
        \path [#1]
        (\tikzinputsegmentfirst) -- (\tikzinputsegmentlast);
      },
    },
  },
  mid arrow/.style={postaction={decorate,decoration={
        markings,
        mark=at position .5 with {\arrow[#1]{stealth}}
      }}},
  start arrow/.style={postaction={decorate,decoration={
        markings,
        mark=at position 0 with {\arrow[#1]{stealth}}
      }}},
  end arrow/.style={postaction={decorate,decoration={
        markings,
        mark=at position .9999 with {\arrow[#1]{stealth}}
      }}},
}

\usetikzlibrary{shapes.misc}
\tikzset{cross/.style={cross out, draw=black, minimum size=2*(#1-\pgflinewidth), inner sep=0pt, outer sep=0pt},
cross/.default={3pt}}

\pgfplotsset{compat=1.16}

\begin{document}
\preprint{AIP/123-QED}

\title[Magnetic Field Design in a Cylindrical High-Permeability Shield]{Magnetic Field Design in a Cylindrical High-Permeability Shield: The Combination of Simple Building Blocks and a Genetic Algorithm}
\author{M.~Packer\textsuperscript{1,+}}
\author{P.~J.~Hobson\textsuperscript{1,+}}%
\author{A.~Davis\textsuperscript{1}}%
\author{N.~Holmes\textsuperscript{1,2}}%
\author{J.~Leggett\textsuperscript{1,2}}%
\author{P.~Glover\textsuperscript{1,2}}%
\author{N.~L.~Hardwicke\textsuperscript{1}}%
\author{M.~J.~Brookes\textsuperscript{1,2}}%
\author{R.~Bowtell\textsuperscript{1,2}}%
\author{T.~M.~Fromhold\textsuperscript{1,×}}%
\affiliation{%
 \textsuperscript{1}School of Physics and Astronomy, University of Nottingham, Nottingham, NG7 2RD, UK. \\
 \textsuperscript{2}Sir Peter Mansfield Imaging Centre, University of Nottingham, Nottingham, NG7 2RD, UK. \\
 \textsuperscript{+}These authors have contributed equally to this work. \\
 \textsuperscript{×}Corresponding author: \href{mailto:Mark.Fromhold@nottingham.ac.uk}{Mark.Fromhold@nottingham.ac.uk}. \\
}
\date{\today}
\begin{abstract}
Magnetically-sensitive experiments and newly-developed quantum technologies with integrated high-permeability magnetic shields require increasing control of their magnetic field environment and reductions in size, weight, power, and cost. However, magnetic fields generated by active components are distorted by high-permeability magnetic shielding, particularly when they are close to the shield's surface. Here, we present an efficient design methodology for creating desired static magnetic field profiles by using discrete coils electromagnetically-coupled to a cylindrical passive magnetic shield. We utilize a modified Green's function solution that accounts for the interior boundary conditions on a closed finite-length high-permeability cylindrical magnetic shield, and determine simplified expressions when a cylindrical coil approaches the interior surface of the shield. We use an analytic formulation of simple discrete building blocks to provide a complete discrete coil basis to generate any physically-attainable magnetic field inside the shield. We then use a genetic algorithm to find optimized discrete coil structures composed of this basis. We use our methodology to generate an improved linear axial gradient field, $\mathrm{d}B_z/\mathrm{d}z$, and transverse bias field, $B_x$. These optimized structures generate the desired fields with less than $1\%$ error in volumes seven and three times greater in spatial extent than equivalent unoptimized standard configurations. This coil design method can be used to optimize active--passive magnetic field shaping systems that are compact and simple to manufacture, enabling accurate control of magnetic field changes in spatially-confined experiments at low cost.
\end{abstract}
\maketitle
\section{Introduction}
The mathematical framework for magnetic field design was first formalized by Romeo and Hoult, who used discrete loops and arcs as the building blocks of a coil basis to generate high-fidelity fields for MRI shimming coils\cite{RandH}. The magnetic field profiles produced by these simple coil building blocks were expanded in a spherical harmonic basis and the harmonic fields related to the geometry, position, and current of the coil basis elements. The geometries were selected, and their positions adjusted, to minimize unwanted signals and, therefore, maximize the fidelity of a desired magnetic field profile. It was subsequently found that inverse methods based on a continuum representation of the current density could allow the design of higher-fidelity magnetic fields, albeit with more computational effort. Pissanetzky first formulated arbitrary current densities on triangular boundary elements\cite{Pissanetzky_1992}, allowing optimal designs to be found through an entirely numerical method. This formulation was later improved upon by Poole\cite{Mpoole}, enabling the flexible design of MRI gradient coils on surfaces of arbitrary geometry using sophisticated 3D-contouring methods. Pseudo-analytical techniques have also been developed on specific surface geometries using Green's function expansions and quadratic optimization methods that enable the rapid design of high-fidelity user-specified magnetic fields in free space\cite{niall1,forbes1,forbes2,forbes3}.

Newly-developed quantum technologies with greater performance and reduced size have further increased the demand for state-of-the-art magnetically-controlled environments. The applications of these technologies range from fundamental physics experiments\cite{Wueaax0800, doi:10.1038/s41598-018-30608-1, SnaddenGradiometer, AIQuantumSensors, MORIC2014287, LiangClock} to biomedical imaging\cite{nature,BOTO2019116099,10.1016/j.neuroimage.2021.118401,TIERNEY2019598,pshwin,doi:10.1002/pd.4976,LEW20172470,doi:10.1161/JAHA.119.013436}. Magnetic field control is required in many of these technologies to trap and manipulate atoms. To translate these laboratory experiments to usable devices in real-world settings, high-permeability passive magnetic shields are used to attenuate stray magnetic fields caused by nearby electronic equipment and/or the local Earth's magnetic field. Specifically, cylindrical and cubic magnetic shields are often used in these systems as they are simple to manufacture, provide good shielding, and can accommodate equipment easily inside them\cite{10.1063/1.1656455, GRABCHIKOV201649, 6217348}. However, previous longstanding methods of magnetic field design do not incorporate the interaction of active current-carrying coils with the high-permeability passive shielding materials. Consequently, if these methods are used to design coils to generate specific magnetic fields in shielded environments, the magnetic shield will distort the field profile, prohibiting the desired level of field control\cite{doi:10.1002/9780470268483.app2}.

Motivated by this problem, several novel numerical and analytical methods have recently been developed that incorporate high-permeability passive shielding material into their design methodologies, to design complex distributions of the current continuum that generate extremely high fidelity magnetic fields inside shielded environments. The numerical methods allow more flexible wire placement whereas the analytical methods allow for physical understanding of the symmetries embedded in the interaction with the shield. The numerical methods use an equipotential scalar field to enforce the boundary condition on the shield's surface, and then account for this in the design of surface currents using boundary elements on arbitrary geometries inside the shield\cite{mkinen2020magneticfield,zetter2020magneticfield}. The analytical methods rely on modifying the Green's function to satisfy the boundary condition on the shield's surface. The currents are then decomposed into an orthogonal basis set where the magnetic fields generated by the combined system can be calculated\cite{PhysRevApplied.14.054004,PhysRevApplied.15.064006}. The current continuum must be carefully discretized into a wire pattern\cite{Hobson2021BespokeMF}, or the high-fidelity fields are not realized physically. Although this error can be estimated analytically for simple current distributions without magnetic shields\cite{Crawford,Nouri}, generally it must be calculated \emph{a posteriori}. This error is determined both by the complexity of representing specific features of the continuum and the response of the magnetic shield to the discretized current. Moreover, although technologies like flex-PCBs\cite{PCBCoils} and 3D-printers\cite{3DPrintingPaper} offer the capability to represent the continuum very precisely, such coils are expensive, time-consuming to manufacture, and hard to repair if there is a breakage. Furthermore, in many of these systems, magnetic field control is not the only area of concern. Optical access, miniaturization, and cost also constrain the development of many of these technologies. In these contexts, using optimally-placed discrete coil designs could allow for simplistic, cost-effective, and accurate generation of magnetic fields with greater optical access. Some simple discrete coil geometries have been formulated that allow the design optimization of magnetic field-generating systems in shielded environments. However, these have been restricted to circular loops and simple transverse fields\cite{Solenoid1,solenoid2,doi:10.1063/1.1719514,LIU2020166846}. Currently, no generalized discrete coil optimization method exists that incorporates the interaction with high-permeability shielding.

Alongside this, multi-objective optimization procedures, such as genetic algorithms, particle swarm optimizations, and differential evolution algorithms, have garnered considerable attention over the past decade because of their ability to find optimal solutions to complicated problems with mixed constraints\cite{542653,MCCALL2005205}. Advances in computational power and code accessibility have made these algorithms much easier to implement\cite{kalyanmoy}. In this paper, using the analytical formulation of cylindrical coils in a cylindrical magnetic shield\cite{PhysRevApplied.14.054004}, the framework of Romeo and Hoult\cite{RandH}, analytic solutions, and a genetic algorithm optimization procedure\cite{NSGA-II}, we present a widely-applicable design methodology that enables the construction of optimized discrete coils in cylindrical magnetic shields. Firstly, we expand on the analytical formulation by determining an approximate form of the magnetic field when the coil is close to the surface of the magnetic shield. Secondly, we formulate a complete coil basis in cylindrical coordinates that allows the simple construction of harmonic fields using discrete coils. Finally, we find optimal configurations of multiple nested sets of the discrete coil basis to generate specified harmonic fields by utilizing a genetic algorithm. By incorporating these different elements, our method enables the simple design of specified magnetic field profiles in high-permeability cylindrical magnetic shields constrained by optical access, cost, and size.

\section{Model}
\begin{figure}[htb]
    \centering
    \includegraphics[width=0.4\columnwidth]{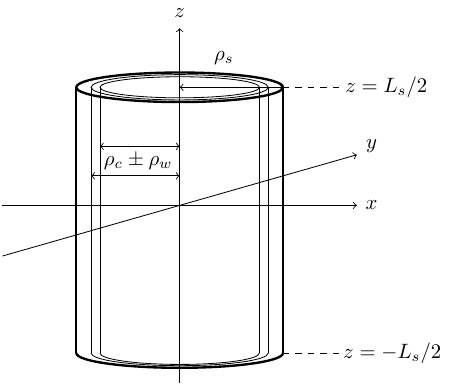}
    \caption{Cylindrical magnetic shield with a high magnetic permeability, $\mu_r\gg1$, of length $L_s$ and inner radius $\rho_s$ with planar end caps located at $z=\pm L_s/2$. A coil of radius $\rho_c$ and equal length to the shield is placed symmetrically inside the shield, and the coils are formed of wire of radius $\rho_w$.}
    \label{fig.magshield}
\end{figure}
Here, we consider a closed high-permeability cylinder of inner radius $\rho_s$ and length $L_s$, with planar end caps located at $z=\pm L_s/2$. Inside this cylinder, a current, $\textbf{J}$, flows on a co-axially nested cylindrical surface of radius $\rho_c$, thickness $2\rho_w$, and length $L_s$, as shown in Fig.~\ref{fig.magshield}, such that $\rho_c+\rho_w\leq\rho_s$. Many magnetic shielding materials, including high-grade mumetal\cite{mu,mu1}, approximate to perfect magnetic conductors, i.e. $\mu_r\to\infty$, under applied fields up to $H=40$~A/m before saturation\cite{muperm}. If the shield is assumed to be a perfect magnetic conductor, the boundary conditions at the shield's surface can be approximated as
\begin{equation}\label{bbound}
    B_\rho\bigg\rvert_{z=\pm L_s/2}=0, \quad
    B_\phi\bigg\rvert_{z=\pm L_s/2,\rho=\rho_s}=0, \quad B_z\bigg\rvert_{\rho=\rho_s}=0.
\end{equation} 
The Green's function solution for the total field, in a region within the cylinder $\rho<\rho_c$, which satisfies \eqref{bbound}, is given by\cite{PhysRevApplied.14.054004}
\begin{align}\label{eq.brdis}
    B_{\rho}\left(\rho,\phi,z\right)=\frac{i\mu_0\rho_c}{2\pi}\sum_{m=-\infty}^{\infty}\sum_{p=-\infty}^{\infty}\int_{-\infty}^{\infty}\mathrm{d}k \ ke^{im\phi}e^{ikz}
     I'_{m}(|k|\rho)R_m(k,\rho_c,\rho_s)J_\phi^{mp}(k),
\end{align}
\begin{align}\label{eq.bpdis}
    B_{\phi}\left(\rho,\phi,z\right)=-\frac{\mu_0\rho_c}{2\pi\rho}\sum_{m=-\infty}^{\infty}\sum_{p=-\infty}^{\infty}\int_{-\infty}^{\infty}\mathrm{d}k \ m\frac{|k|}{k}e^{im\phi}e^{ikz}
     I_{m}(|k|\rho) R_m(k,\rho_c,\rho_s)J_\phi^{mp}(k),
\end{align}
\begin{align}\label{eq.bzdis}
    B_{z}\left(\rho,\phi,z\right)=-\frac{\mu_0\rho_c}{2\pi}\sum_{m=-\infty}^{\infty}\sum_{p=-\infty}^{\infty}\int_{-\infty}^{\infty}\mathrm{d}k \ |k|e^{im\phi}e^{ikz}
     I_{m}(|k|\rho)R_m(k,\rho_c,\rho_s)J_\phi^{mp}(k),
\end{align}
where $R_m(k,\rho_c,\rho_s)=K'_{m}(|k|\rho_c)-I'_{m}(|k|\rho_c)K_{m}(|k|\rho_s)/I_{m}(|k|\rho_s)$, and $J_\phi^{mp}(k)$ is the Fourier transform with respect to $z$ and $\phi$ of the $p^{\textnormal{th}}$ reflected image current determined via the method of mirror images\cite{jackson}, where the $p=0$ term represents the Fourier transform of the actual current distribution which is confined to the region $|z'|<L_s/2$
\begin{equation}\label{eq.fp}
    J_\phi^{mp}(k)=\frac{1}{2\pi}\int_{0}^{2\pi}\mathrm{d}\phi' \ e^{-im\phi'}\int_{-\infty}^{\infty}\mathrm{d}z'\ e^{-ikz'}J_\phi\left(\phi',(-1)^p\left(z'+pL_s\right)\right),
\end{equation}
where ($\phi'$,$z'$) specify position on the current-carrying surface.

\begin{figure}[htb]
    \centering
    \includegraphics[width=0.8\columnwidth]{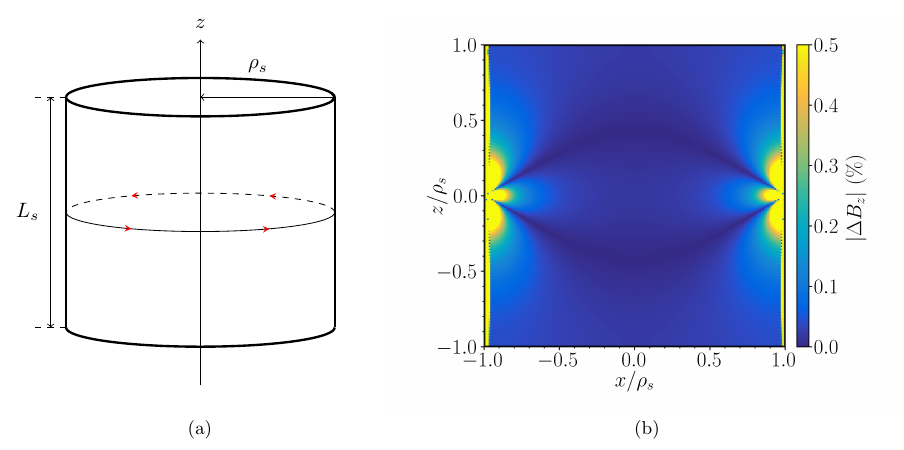}
    \caption{(a) Schematic diagram of a loop of radius $0.99\rho_s$ at position $z=0$ in a closed magnetic shield of radius $\rho_s$ and length $L_s=2\rho_s$. (b) Color map showing the absolute error, $|{\Delta}B_z|$, between the axial zeroth-order contribution in \eqref{eq.bzdisbig} and the exact solution in \eqref{eq.bzdis} for the example depicted in (a).}
    \label{fig.error_zc}
\end{figure}

Here, we use this formulation to design simple discrete coils inside of closed cylindrical magnetic shields. To maximize the available interior volume of the system, as is normally required for real-world applications, we consider discrete coils that are positioned at the shield's inner surface, $\rho_c=\rho_s-\rho_w$, and determine their parameters using forward numerical optimization techniques, thereby circumventing discretization error entirely. When the coil is pressed against the inner surface of the shield, we can expand the magnetic field as a power series of the (small) wire radius such that
\begin{equation}
    \mathbf{B}\left(\rho,\phi,z\right)=\mathbf{B}^0\left(\rho,\phi,z\right)+\rho_w\mathbf{B}^1\left(\rho,\phi,z\right)+\rho_w^2\mathbf{B}^2\left(\rho,\phi,z\right)+ \hdots,
\end{equation}
where the $\mathbf{B}^\nu$ terms are $\nu^{\textnormal{th}}$ order field perturbations for $\nu\in\mathbb{Z}^{0+}$. If the radius of wire is sufficiently small compared to the radius of the magnetic shield, the magnetic field can be approximated while only introducing small deviations. Here, we give the example of a simple loop placed at the center of a shield with aspect ratio $L_s/(2\rho_s)=1$ and wire radius $\rho_w=0.01\rho_s$. The error between the zeroth-order term and the complete solution is less than $0.016\%$ at the center, as shown by Fig.~\ref{fig.error_zc}, but moving towards the cylindrical wall it increases. Discounting the region close to the shield, the error within radial position $\rho<0.8\rho_s$ is less than $0.25\%$. Henceforth, in this paper, we assume that $\rho_w<0.01\rho_s$ and use only the zeroth-order term to design coils in this regime. The magnetic field components are simplified using the Wronskian, resulting in the governing equations
\begin{align}\label{eq.brdisbig}
    B_{\rho}\left(\rho,\phi,z\right)=-\frac{i\mu_0}{2\pi}\sum_{m=-\infty}^{\infty}\sum_{p=-\infty}^{\infty}\int_{-\infty}^{\infty}\mathrm{d}k \ \frac{k}{|k|}e^{im\phi}e^{ikz}
     \frac{I'_{m}(|k|\rho)}{I_{m}(|k|\rho_s)}J_\phi^{mp}(k),
\end{align}
\begin{align}\label{eq.bpdisbig}
    B_{\phi}\left(\rho,\phi,z\right)=\frac{\mu_0}{2\pi\rho}\sum_{m=-\infty}^{\infty}\sum_{p=-\infty}^{\infty}\int_{-\infty}^{\infty}\mathrm{d}k \ \frac{m}{k}e^{im\phi}e^{ikz}
     \frac{I_{m}(|k|\rho)}{I_{m}(|k|\rho_s)}J_\phi^{mp}(k),
\end{align}
\begin{align}\label{eq.bzdisbig}
    B_{z}\left(\rho,\phi,z\right)=\frac{\mu_0}{2\pi}\sum_{m=-\infty}^{\infty}\sum_{p=-\infty}^{\infty}\int_{-\infty}^{\infty}\mathrm{d}k \ e^{im\phi}e^{ikz}
     \frac{I_{m}(|k|\rho)}{I_{m}(|k|\rho_s)}J_\phi^{mp}(k).
\end{align}
For a setup where the $\rho_w>0.01\rho_s$, the validity of the approximation should be determined for each individual scenario and adjusted appropriately for a given field design tolerance and experimental system.

\section{Coil Basis}
In free space, the magnetic field can be represented as the gradient of a scalar potential, $\mathbf{B}=-\nabla\Psi$. The scalar potential and magnetic field, \eqref{eq.brdisbig}-\eqref{eq.bzdisbig}, both satisfy Laplace's equation. Following Romeo and Hoult\cite{RandH}, we express the magnetic field as the set of real spherical harmonics in spherical polar coordinates,
\begin{equation}\label{eq.sh}
    \mathbf{B}(r,\theta,\phi)=\boldsymbol{\nabla}\sum_{n=0}^{\infty} \sum_{m=-n}^{n} C_{n,m}r^{n}P_{n,|m|}\left(\cos\theta\right)
    \begin{pmatrix}
    \cos\left(|m|\phi\right)\\
    \sin\left(|m|\phi\right)
    \end{pmatrix},
    \qquad
    \begin{matrix}
    m\geq0\\
    m<0
\end{matrix}
\end{equation}
where the harmonic fields are classified through their order, $n$, and degree, $m$. Each harmonic has a magnitude, $C_{n,m}$, and a $\theta$ dependence that is described by one of the Ferrer’s associated Legendre polynomials, $P_{n,|m|}\left(\cos\theta\right)$. The degree is divided into two cases, $m=0$ and $|m|>0$. The $m=0$ harmonic fields exhibit total azimuthal symmetry and are known as zonal harmonics, $Z_{n}$. The $|m|>0$ harmonic fields exhibit $m$-fold azimuthal symmetry and are known as tesseral harmonics, $T_{n,m}$, where negative $m<0$ harmonic fields are $\pi/(2|m|)$ azimuthal rotations of their positive $m>0$ counterparts.

In appendix~A, we solve for the axial magnetic field component from~\eqref{eq.sh},
\begin{equation}\label{eq.shbz}
    B_z(r,\theta,\phi)=\sum_{n=1}^\infty\sum_{m=-n+1}^{n-1}\ C_{n,m}(n+|m|)r^{n-1}P_{n-1,|m|}\left(\cos\theta\right)\begin{pmatrix}
    \cos\left(|m|\phi\right)\\
    \sin\left(|m|\phi\right)
    \end{pmatrix}.
    \qquad
    \begin{matrix}
    m\geq0\\
    m<0
    \end{matrix}
\end{equation}
Due to the symmetry of the associated Legendre polynomials, the parity of the axial field is even if $n+m=2\nu+1$ and odd if $n+m=2\nu$, respectively, for $\nu\in\mathbb{Z}$. No axial field exists where $n=|m|$. Although the complete set of harmonic fields do not exist within the axial field, any harmonic can be indirectly selected using it due to its relationship to the scalar potential. Other magnetic fields may be used to select harmonics, however their functional forms are more complex\cite{RandH}. Thus, the axial field is most appropriate for construction of any magnetic field provided the correct current density basis is chosen such that the harmonics that are not present in the axial field can be removed independently. 

\begin{figure}[!htb]
    \centering
    \includegraphics[width=0.6\columnwidth]{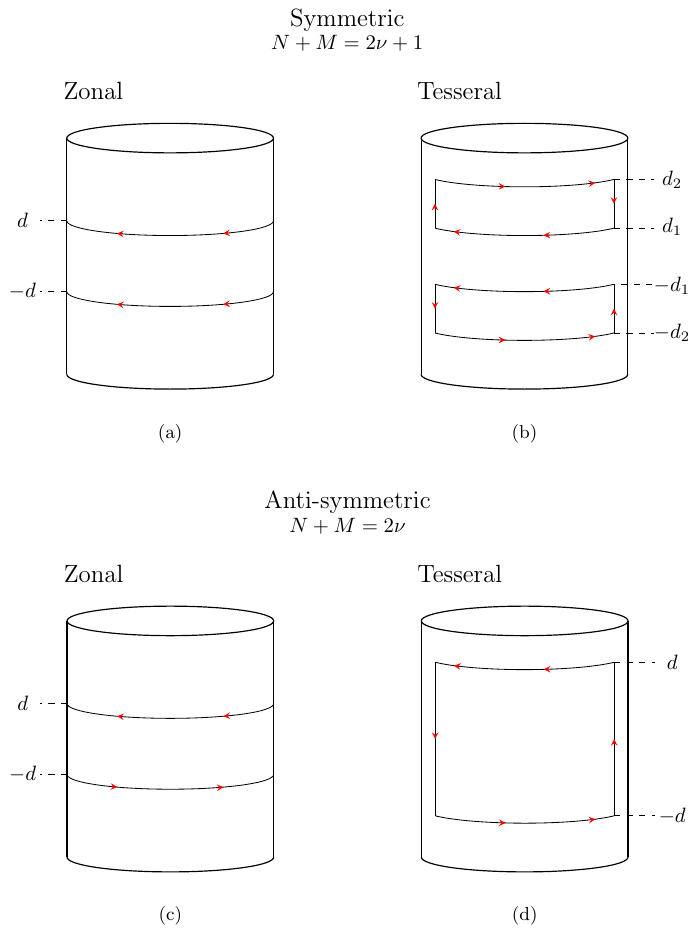}
    \caption{Azimuthal and axial variation in the basis currents on the ${\phi}z$-plane required by \eqref{eq.jphias} to generate (a)-(b) symmetric, $(N+M)=2\nu+1$, and (c)-(d) anti-symmetric, $(N+M)=2\nu$, zonal and tesseral harmonics, respectively, where $N$ and $M$ are the order and degree of the harmonic and $\nu\in\mathbb{Z}$. Red arrow heads show the direction of current flow.}
    \label{fig.symanti}
\end{figure}

\begin{figure}[!htb]
    \centering
    \includegraphics[width=0.8\columnwidth]{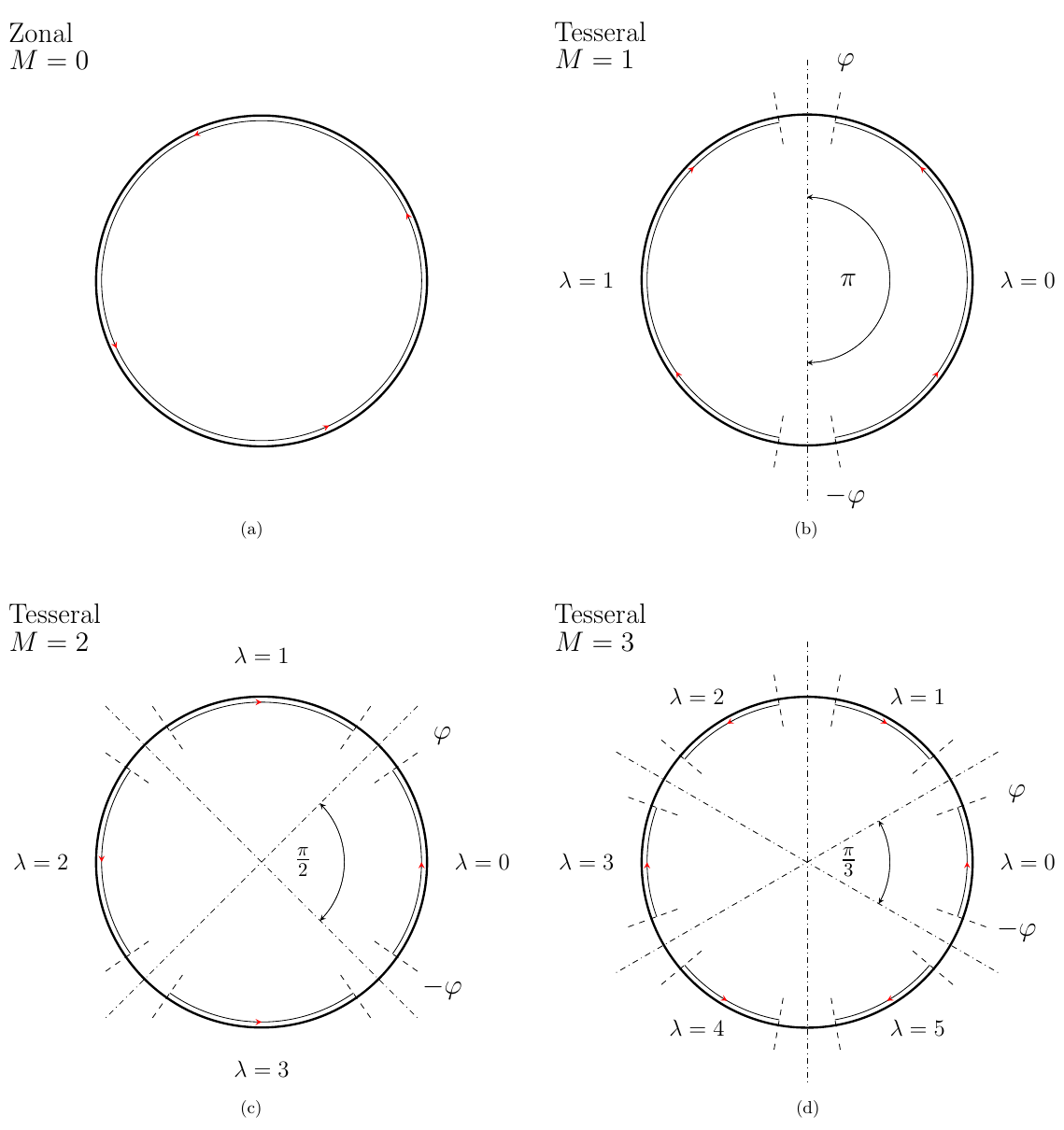}
    \caption{Azimuthal variation in the basis currents on the $\rho\phi$-plane required by \eqref{eq.azvar} to generate (a) the zonal, $M=0$, and (b-d) tesseral harmonics of degree one, two, and three, $M=(1-3)$, respectively, where the azimuthal arc length for each period, $\lambda$, is given by $2\varphi$. Red arrow heads show the direction of current flow.}
    \label{fig.arcs}
\end{figure}

The axial magnetic field, \eqref{eq.bzdis}, is directly related to the Fourier transform of the azimuthal current density. To design coils effectively using the axial field, the axial parity and azimuthal symmetry of the azimuthal current density must enable $\phi$ and $z$ variations to be decoupled independently. To generate zonal harmonics, this requires closed circular azimuthal current loops with complete azimuthal symmetry. To generate tesseral harmonics, this requires a set of arcs of the same azimuthal periodicity as the desired harmonic. Because arcs are not continuous, they must be linked via axial connections, forming saddle-like systems\cite{Saddle}. Due to the preserved symmetries of the Legendre polynomials in the axial field, \eqref{eq.shbz}, pairs of axially separated coils, centered about the origin of the shield, with symmetric or anti-symmetric current flows can only generate odd and even parity harmonics, respectively. From this parity, the symmetry of a specific order of harmonic, $n=N$, and, subsequently, axial coil symmetry, can then be chosen to select the required field symmetries within the system. Thus, there are four units which form the building blocks of the coil basis which will be used to construct any arbitrary harmonic field using the axial field -- symmetric and anti-symmetric loops and arcs -- as shown in Fig.~\ref{fig.symanti}. To formulate these mathematically, let us decompose the current density into axial and azimuthal components
\begin{equation}\label{eq.jphias}
    J_\phi(\phi',z')=I\Phi(\phi')Z(z'),
\end{equation}
where $I$ is the current in the wire. The axial variation of a symmetric or anti-symmetric pair at axial positions $z'=d$ and $z'=-d$, respectively, is given by
\begin{equation}\label{eq.das}
    Z^{\pm}(z')=\delta(z'-d)\pm\delta(z'+d),
\end{equation}
with the resulting $p^{\textnormal{th}}$ reflected Fourier transform from \eqref{eq.fp} written as
\begin{equation}\label{eq.fppair}
    J_\phi^{mp}(k)=e^{ikpL_s}\left(e^{-(-1)^pikd}\pm e^{(-1)^pikd}\right)\Phi^m,
\end{equation}
where the azimuthal Fourier transform of the azimuthal variation of the current density is
\begin{equation}\label{eq.axft}
    \Phi^m=\frac{1}{2\pi}\int_{0}^{2\pi}\mathrm{d}\phi' \ e^{-im\phi'}\Phi\left(\phi'\right).
\end{equation}
To maximize a specific degree of harmonic, $m=M$, with either complete azimuthal symmetry, $M=0$, or periodicity, $\pi/|M|$, the azimuthal component is chosen according to the desired harmonic field, given by
\begin{equation}\label{eq.azvar}
    \Phi(\phi')=
    \begin{cases}
     \sum_{\lambda=0}^{2M-1} (-1)^{\lambda}\left[ H\left(\phi'+\varphi-\frac{\lambda\pi}{M}\right)-H\left(\phi'-\varphi-\frac{\lambda\pi}{M}\right)\right], & M>0 \\ 
     1, & M=0\\ 
     \sum_{\lambda=0}^{2|M|-1} (-1)^{\lambda}\left[ H\left(\phi'+\varphi-\frac{\lambda\pi}{|M|}-\frac{\pi}{2|M|}\right)-H\left(\phi'-\varphi-\frac{\lambda\pi}{|M|}-\frac{\pi}{2|M|}\right)\right], & M<0
     \end{cases}
\end{equation}
where $H(x)$ is the Heaviside function. These azimuthal variations are illustrated in Fig.~\ref{fig.arcs}. The azimuthal Fourier transform, from \eqref{eq.axft}, is then found to be
\begin{equation} \label{eq.azft}
    \Phi^m(\varphi)=\begin{cases}
    \frac{\sin(m\varphi)}{\pi m}\sum_{\lambda=0}^{2M-1}(-1)^{\lambda}e^{-\frac{im\lambda\pi}{M}}, & M>0 \\
    \delta_{m0}, & M=0\\
    \frac{\sin(m\varphi)}{\pi m}e^{-\frac{im\pi}{2|M|}}\sum_{\lambda=0}^{2|M|-1}(-1)^{\lambda}e^{-\frac{im\lambda\pi}{|M|}}. & M<0
    \end{cases}
\end{equation}

Substituting \eqref{eq.fppair} into \eqref{eq.bzdisbig}, and noting that the expression can be written in terms of a Fourier series, the axial magnetic field generated by symmetric and anti-symmetric pairs is given by
\begin{equation}\label{eq.bztotal}
    B^{\pm}_{z}\left(\rho,\phi,z\right)=\frac{2\mu_0I}{L_s}\sum_{m=-\infty}^{\infty}\  b_m^{\pm}(\rho,z;d)e^{im\phi}\Phi^m,
\end{equation}
where
\begin{align}\label{eq.bzsym}
    b_m^{+}(\rho,z;d)=\sum_{p\textnormal{ even}} \ \cos\left(\frac{\pi pz}{L_s}\right)\cos\left(\frac{\pi pd}{L_s}\right)
    \frac{I_{m}\left(\left|\frac{\pi p}{L_s}\right|\rho\right)}{I_m\left(\left|\frac{\pi p}{L_s}\right|\rho_s\right)},
\end{align}
\begin{align}\label{eq.bzanti}
    b_m^{-}(\rho,z;d)=\sum_{p\textnormal{ odd}} \ \sin\left(\frac{\pi pz}{L_s}\right)\sin\left(\frac{\pi pd}{L_s}\right)
    \frac{I_{m}\left(\left|\frac{\pi p}{L_s}\right|\rho\right)}{I_m\left(\left|\frac{\pi p}{L_s}\right|\rho_s\right)},
\end{align}
are symmetric and anti-symmetric axial magnetic field variations, respectively, of the coil basis for $p\in\mathbb{Z}$. Using this coil basis that generates zonal and tesseral, symmetric and anti-symmetric fields we may now begin to construct coil structures that select specific harmonic fields.

Alternatively, a spherical coil basis\cite{RandH} may be used with loops at different zenith angles and axial positions to construct the complete set of harmonic fields. However, rotated zonal loops do not sit exactly on the interior surface of the magnetic shield unless they are projected onto ellipses, for which exact solutions are hard to generate.

\section{Harmonic Selection}
\begin{figure}[!h]
    \centering
    \includegraphics[width=0.6\columnwidth]{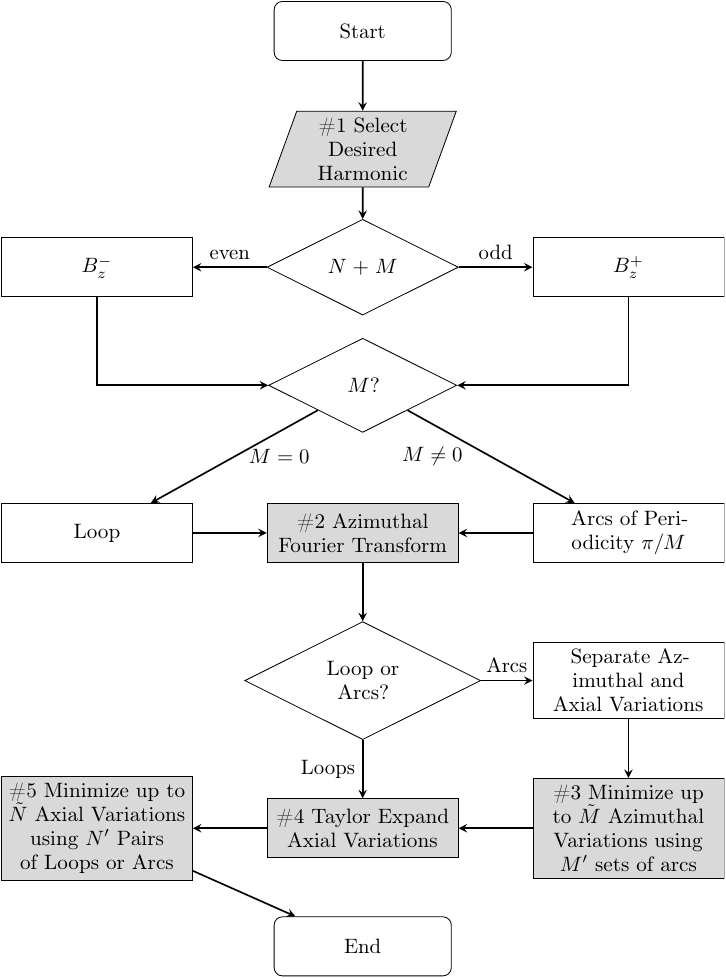}
    \caption{Flow diagram describing the harmonic selection process for generating a desired harmonic of order $N$ and degree $M$ using $N'$ axial pairs of loops or arcs with $M'$ arcs at each axial position. The steps which we follow in the main text are highlighted in grey. Step~\#3 is skipped when $M=0$.}
    \label{fig.flow}
\end{figure}
We now propose a methodology for designing a coil to generate a specific spherical harmonic variation in any vector direction using the coil basis. The road map of this harmonic selection process is presented in Fig.~\ref{fig.flow}. 

\begin{figure}
    \centering
    \includegraphics[width=0.8\columnwidth]{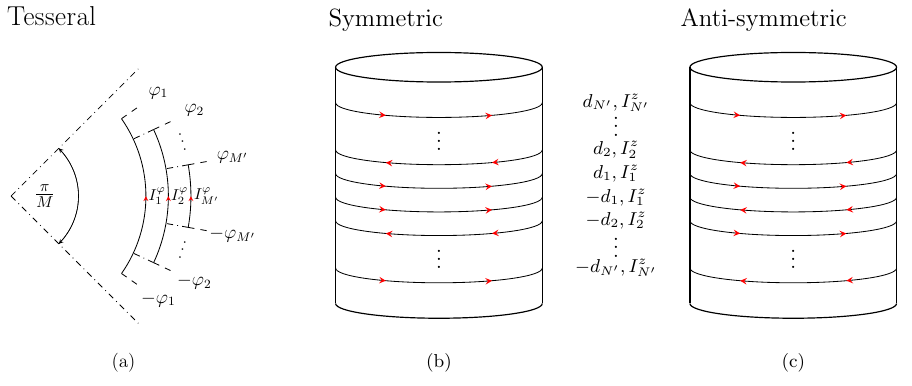}
    \caption{Sets of the basis currents from \eqref{eq.jphias} to generate higher fidelity representations of (a) tesseral harmonics of degree $M$ presented on the $\rho\phi$-plane and (b)-(c) symmetric and anti-symmetric zonal harmonics, respectively, presented on the ${\phi}{z}$-plane, where $N'$ and $M'$ are the number of basis currents used. Red arrow heads show the direction of current flow.}
    \label{fig.multiarcs}
\end{figure}

First, we select a desired magnetic field harmonic of order $N$ and degree $M$ (Step~\#1). The azimuthal variations and, subsequently, the degrees of the harmonics generated, as described in \eqref{eq.bztotal}, are determined by the periodicity of a given coil configuration. Thus, to maximize the degree of any desired harmonic we must consider the azimuthal Fourier transform, \eqref{eq.azft} (Step~\#2). For $M=0$, it is apparent that loops only generate fields of degree $m=0$ and, so, do not require azimuthal optimization. For $|M|>0$, however, sets of arcs of periodicity $\pi/|M|$ generate an infinite number of harmonic fields of degree $m=(2\nu+1)M$, where $\nu \in \mathbb{Z}^{0+}$. Therefore, to maximize the desired degree of a tesseral harmonic field, the angular length, $\varphi$, should be adjusted to eliminate as many undesired azimuthal variations as possible. From analysis of \eqref{eq.azft}, the leading-order error term of degree $m=3M$ is removed if
\begin{equation}\label{eq.arc1}
    \sin(3M\varphi)=0.
\end{equation}
However, depending on the required accuracy of the desired field, further variations might need to be removed. To achieve this, additional arcs of angular length $\varphi_j$ and azimuthal turn ratios, $I_j^\varphi$, can be used to allow multiple degrees to be minimized simultaneously, as shown in Fig.~\ref{fig.multiarcs}a. Hence, generalizing \eqref{eq.arc1}, we can use $M'$ arcs simultaneously to minimize $\tilde{M}$ degrees of harmonics (Step~\#3),
\begin{equation}\label{eq.phimin}
    \min_{\varphi_j,I^\varphi_j}\left[\sum_{j=1}^{M'}I^\varphi_j\sin((2\nu+1)M\varphi_j)\right], \qquad \nu\in \mathbb{Z}:\nu\in[1,\tilde{M}].
\end{equation}
The harmonics in \eqref{eq.phimin} can be nulled completely for simple integer $I_j^\varphi$ by substituting the appropriate Chebyshev polynomials or, easily and quickly in many cases, by using commercial root-finding software. For practical applications, $I_j^\varphi$ must be integer ratios of one-another and connected in series, limiting the space in which optimal $\varphi_j$ can exist. Typically, the best solutions have significant angular lengths and azimuthal turn ratios within an order of magnitude of each other to prevent the finite size of the wires from introducing unwanted deviations from the desired field. It should also be noted that designs with counter-propagating current flows, i.e. both positive and negative $I_j^\varphi$, are useful if there are specific regions where wires are prohibited, providing additional flexibility when designing coil setups, but such designs may be very power inefficient. In extreme cases where $\tilde{M}$ is large and/or the angular lengths are highly restricted, a multi-variate optimization algorithm, as described in section~V, may be employed to solve for multiple $\varphi_j$ and $I_j^\varphi$ to minimize \eqref{eq.phimin}.

The same logic can also be applied to the radial and axial field variations to remove harmonics of odd or even parity. To illustrate this, we first transform the spherical harmonic axial field, \eqref{eq.shbz}, into cylindrical coordinates, and separate it into terms of even and odd parity,
\begin{align}\label{eq.shbz2}
    B_z=
    &\Bigg[ \sum_{n=0}^\infty\sum_{m=-\infty}^{\infty}\ C_{2n+|m|+1,m}(2n+2|m|+1)(\rho^2+z^2)^\frac{2n+|m|}{2} P_{2n+|m|,|m|}\left(\frac{z}{(\rho^2+z^2)^{1/2}}\right) \ + \\
    &\sum_{n=1}^\infty\sum_{m=-\infty}^{\infty}\ C_{2n+|m|,m}(2n+2|m|)(\rho^2+z^2)^\frac{2n+|m|-1}{2}P_{2n+|m|-1,|m|}\left(\frac{z}{(\rho^2+z^2)^{1/2}}\right) \Bigg], \begin{pmatrix}\cos\left(|m|\phi\right) \\ \sin\left(|m|\phi\right) \end{pmatrix} \qquad \begin{matrix} m\geq0\\m<0 \end{matrix} \qquad \nonumber.
\end{align}
After analyzing \eqref{eq.shbz2} and \eqref{eq.bztotal}, we can see that the radial and axial dependence of every harmonic of order $n$ and degree $m$, excluding $m\neq n$, must be completely contained within the symmetric and anti-symmetric axial field variations, \eqref{eq.bzsym}-\eqref{eq.bzanti}. Therefore, we can write these variations as
\begin{equation}\label{eq.bzsymex}
    b_m^{+}(\rho,z;d)=\sum_{n=0}^\infty
    \tilde{C}_{2n+|m|+1,m}(d,\rho_s,L_s)(\rho^2+z^2)^\frac{2n+|m|}{2}P_{2n+|m|,|m|}\left(\frac{z}{(\rho^2+z^2)^{1/2}}\right),
\end{equation}
\begin{equation}\label{eq.bzasymex}
    b_m^{-}(\rho,z;d)=\sum_{n=1}^\infty
    \tilde{C}_{2n+|m|,m}(d,\rho_s,L_s)(\rho^2+z^2)^\frac{2n+|m|-1}{2}P_{2n+|m|-1,|m|}\left(\frac{z}{(\rho^2+z^2)^{1/2}}\right),
\end{equation}
where $\tilde{C}_{n,m}(d,\rho_s,L_s)$ are effective harmonic magnitudes, which depend only on the coil and shield parameters. To derive $\tilde{C}_{n,m}(d,\rho_s,L_s)$, we substitute Taylor expansions of trigonometric and Bessel functions into \eqref{eq.bzsym}-\eqref{eq.bzanti} and group the spatial variations into their constituent spherical harmonic functions (Step~\#4). This must be done on a case-by-case basis since only specific sets of harmonics exist within each of the basis coils.

Assuming that the required azimuthal variations are eliminated using \eqref{eq.phimin}, the remaining undesired harmonics constitute $\tilde{N}$ different axial variations of degree $M$ in the axial field. To generate a desired harmonic, we minimize these variations by simultaneously optimizing $N'$ pairs of loops/arcs at positions $d_i$ with axial turn ratios $I_i^z$ (Step~\#5), for the symmetric case
\begin{equation}\label{eq.zminsym}
    \min_{I^z_i, d_i}\left( \sum_{i=1}^{N'} I^z_i \tilde{C}_{2n+M+1,M}(d_i,\rho_s,L_s) \right), \qquad n\in \mathbb{Z}:n\in \left( [0,\tilde{N}] \ \mathrm{ excluding } \ (N=2n+M+1) \right),
\end{equation}
and the anti-symmetric case
\begin{equation}\label{eq.zminanti}
    \min_{I^z_i, d_i}\left( \sum_{i=1}^{N'} I^z_i \tilde{C}_{2n+M,M}(d_i,\rho_s,L_s) \right), \qquad n \in \mathbb{Z}:n\in \left( [1,\tilde{N}+1] \ \mathrm{ excluding } \ (N=2n+M) \right).
\end{equation}
This is illustrated for zonal symmetric and anti-symmetric loops in Fig.~\ref{fig.multiarcs}b-c. Note that we exclude the desired order, where $N=2n+M+1$ and $N=2n+M$ in the symmetric and anti-symmetric cases, respectively, from the minimization. Depending on the use-case, conditions that $I^z_i$ is an integer with a limited magnitude may constrain the optimization landscape. As Step~\#4 needs to be applied on a case-by-case basis, we shall now demonstrate the harmonic selection process with a simple example.

\subsection*{Example: Zonal linear axial gradient}
\begin{figure}[htb]
    \centering
    \includegraphics[width=0.8\columnwidth]{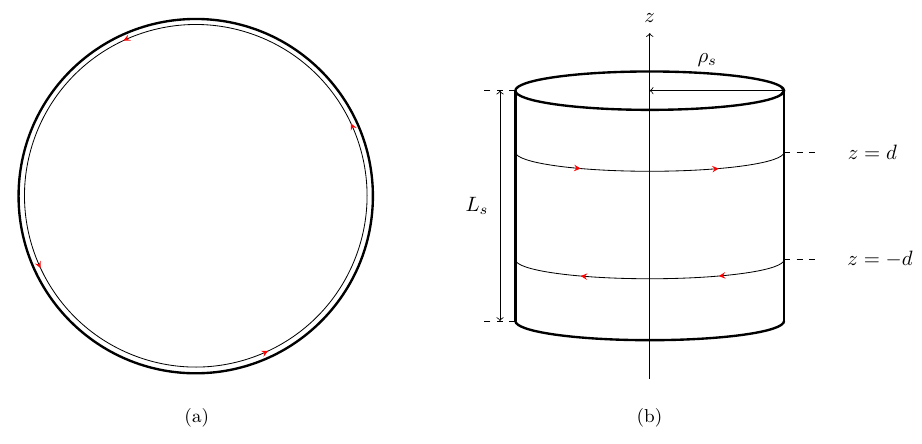}
    \caption{Schematic diagram of an anti-symmetric pair of current-carrying loops of radius $\rho_s$ showing (a) their azimuthal position (thin black circle with red arrow heads indicating current flow direction) within the magnetic shield (thick circle) and (b) their axial positions at $z={\pm}d$ placed symmetrically from the axial center at $z=0$ of a closed magnetic shield of radius $\rho_s$ and length $L_s=2\rho_s$.}
    \label{fig.optahelmpic}
\end{figure}

\begin{figure}[htb]
\centering
\includegraphics[width=0.8\columnwidth]{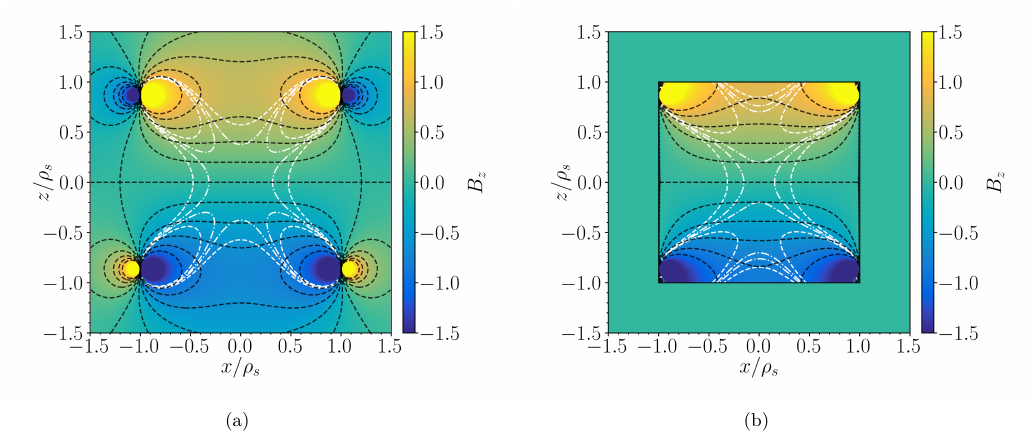}
\caption{Color maps showing the magnitude of the normalized axial magnetic field, $B_z$, in the $xz$-plane generated by the coil depicted in Fig.~\ref{fig.optahelmpic} in the anti-Helmholtz arrangement with separation, $d=\pm\left(\sqrt{3}/2\right)\rho_s$ in two situations (a) in free space and (b) placed symmetrically around the origin of a closed magnetic shield of radius $\rho_s$ and length $L_s=2\rho_s$ (solid black outline). White contours enclose the regions where the gradient of the normalized axial field with respect to $z$ deviates from unity (i.e. a perfectly uniform axial field gradient) by less than $5$\% (dashed curves) and less than $1$\% (dot-dashed curves). Black contours represent lines of constant magnetic flux (dashed curves). The resistance, field per unit current, and inductance of the coil both in free space and inside a unit length magnetic shield are presented in Table~\ref{table.stats}.
}
\label{fig.inshieldhomo}
\end{figure}

An anti-Helmholtz pair within the bore of a cylindrical high-permeability cylinder is presented in Fig.~\ref{fig.optahelmpic}. The anti-Helmholtz configuration uses a pair of anti-symmetric axial loops to generate a scalar harmonic field, $Z_{2}$, which produces an axial linear gradient with respect to axial position, $\mathrm{d}B_z/\mathrm{d}z$. The optimal loop position in free space, $d=\left(\sqrt{3}/2\right)\rho_c$, can be derived by eliminating the cubic variations in the generated field, i.e. the $Z_4$ scalar harmonic\cite{RandH}. In Fig.~\ref{fig.inshieldhomo}a and Fig.~\ref{fig.inshieldhomo}b we examine the field linearity of the anti-Helmholtz coils located, respectively, in free space and within a cylindrical magnetic shield of aspect ratio $L_s/(2\rho_s)=1$. The presence of the magnetic shield affects the inductance of the coils and the profile of the field that they generate. In particular, coupling to the shield increases the inductance of the coils and the field gradient that they generate by approximately a factor of two. In addition, the magnetic shield amplifies the non-zero cubic variations in the field profile, causing it to deviate from a linear variation and reducing, by a factor of approximately three, the volume (bounded by dot-dashed curves in Fig.~\ref{fig.inshieldhomo}) wherein the generated and desired field gradient are within $1\%$ of one another. Evidently, new optimal coil separations must be determined to improve the accuracy of the magnetic field gradient in shielded environments.

\begin{figure}[htb]
    \centering
    \includegraphics[width=0.8\columnwidth]{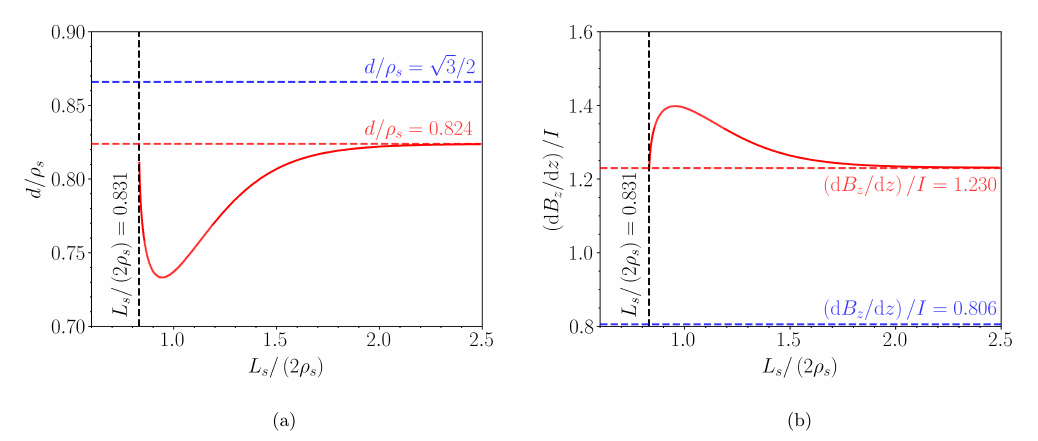}
    \caption{(a) Optimal normalized separation, $d/\rho_s$, of the anti-symmetric pair, depicted in Fig.~\ref{fig.optahelmpic}, to generate the zonal $Z_2$ harmonic as the length of the shield increases (red curve). Horizontal dashed lines (red and blue) show the analytical values of $d=0.824\rho_s$ and $d=\left(\sqrt{3}/2\right)\rho_s$ obtained in the long shield limit ($L_s\gg2\rho_s$) and in free space, respectively. (b) Gradient per current, $(\mathrm{d}B_z/\mathrm{d}z)/I$, of the optimal anti-symmetric pair as the length of the shield increases (red curve). Horizontal dashed lines (red and blue) show the values of $\mathrm{d}B_z/\mathrm{d}z=1.230I$ and $\mathrm{d}B_z/\mathrm{d}z=0.806I$ obtained in the long shield limit and in free space, respectively. Vertical dashed line (black) in (a) and (b) shows the minimum shield length $L_s=1.62\rho_s$ below which no optimal solution can be found.}
    \label{fig.optahelm}
\end{figure}

The axial magnetic field generated by a pair of loops with counter-flowing currents and located co-axially on the interior surface of the high-permeability shield is, from \eqref{eq.fppair}, \eqref{eq.azft}, and \eqref{eq.bztotal},
\begin{align}\label{eq.bznewhomo}
    B_{z}\left(\rho,\phi,z\right)=\frac{2\mu_0I}{L_s}b^-_0(\rho,z;d).
\end{align}
As explained in the previous section, any given magnetic profile in the system can be found by expanding the spatially-varying functions. Using \eqref{eq.bzanti}, and substituting the well-known series expansions,
\begin{equation}
    I_m(x)=\sum_{l=0}^\infty \frac{1}{l!(l+m)!}\left(\frac{x}{2}\right)^{2l+m} \qquad \textnormal{and} \qquad \sin(x)=\sum_{l=0}^\infty\frac{(-1)^lx^{2l+1}}{(2l+1)!},
\end{equation}
the axial field generated by the anti-symmetric pair, \eqref{eq.bznewhomo}, can be written in terms of the harmonic fields
\begin{equation}
    B_{z}\left(\rho,\phi,z\right)=\frac{2\mu_0I}{L_s}\Bigg({\pi}z\tilde{C}_{2,0}\left(d,\rho_s,L_s\right)+{\pi}^3\left(\frac{z\rho^2}{4}-\frac{z^3}{6}\right)\tilde{C}_{4,0}\left(d,\rho_s,L_s\right)+\hdots\Bigg),
\end{equation}
where the effective harmonic magnitudes are given by
\begin{align}\label{eq.iBzg}
    \tilde{C}_{2n,0}\left(d,\rho_s,L_s\right)=\frac{1}{L_s^{2n-1}}\sum_{p=1}^\infty (2p-1)^{2n-1}\frac{\sin\left(\frac{\pi d(2p-1)}{L_s}\right)}{I_{0}\left(\frac{\pi(2p-1)\rho_s}{L_s}\right)}.
\end{align}
Using \eqref{eq.iBzg}, the optimal positions, $z={\pm}d$, of the coils in an anti-symmetric pair can be determined so that the leading-order axial variation in the desired field is removed when the coils are enclosed by a shield with a given aspect ratio,
\begin{equation}
    \tilde{C}_{4,0}\left(d,\rho_s,L_s\right)=\frac{1}{L_s^3}\sum_{p=1}^\infty (2p-1)^{3}\frac{\sin\left(\frac{\pi d(2p-1)}{L_s}\right)}{I_{0}\left(\frac{\pi(2p-1)\rho_s}{L_s}\right)}=0.
\end{equation}
In Fig.~\ref{fig.optahelm}a, we show the optimal separation calculated versus the shield aspect ratio by an exhaustive numerical search. Figure~\ref{fig.optahelm}b shows the corresponding variation of the gradient per unit current. The red dotted lines in Fig.~\ref{fig.optahelm}a and Fig.~\ref{fig.optahelm}b show, respectively, the optimal separation, $d=0.824\rho_s$, in the limit that the shield aspect ratio tends to infinity, and its corresponding gradient per unit current, $\mathrm{d}B_z/\mathrm{d}z=1.230I$. The blue dotted lines in Fig.~\ref{fig.optahelm}a and Fig.~\ref{fig.optahelm}b are, respectively, the coil separation for the standard anti-Helmholtz configuration, $d=\sqrt{3}\rho_s/2$, and its gradient per unit current, $\mathrm{d}B_z/\mathrm{d}z=0.806I$, that is generated in free space. Due to the interaction and finite length of the magnetic shield there exists a shield aspect ratio, $0<L_s/(2\rho_s)\lessapprox0.831$, where no coil separation entirely removes the cubic variation in the field. In this case, to determine the optimal separation, contributions from both the cubic and quintic variations should be minimized, but not nulled entirely, to achieve the most uniform field linearity for a given application. However, minimization of further variations becomes more difficult since the effective harmonic magnitudes become increasingly sensitive to the precise values of $d_i$, $\rho_s$, and $L_s$.

\section{Genetic Algorithm Optimization}
To simultaneously solve for multiple axial variations in a computationally efficient manner we use a genetic algorithm. The optimal continuous separations, $d_i$, and discrete turn ratios, $I^z_i$, for $N'$ loops or arcs are found by minimizing the amplitudes of a set of undesired harmonic fields. We formulate the optimization problem by using the set of arbitrary user-defined undesired harmonic fields of order $n \in \mathbb{Z}:n \in [\tilde{n}_1,\tilde{n}_{\tilde{N}}]$ and degree $M$ as the objective functions:
\begin{equation}\label{eq.gaobj}
    \begin{cases}
     \min f_1 = \tilde{C}_{\tilde{n}_1,M}\left(\rho_s,L_s;d_1,\hdots,d_{N'},I_1,\hdots,I_{N'}\right), \\
     \hspace{100pt}\vdots \\
     \min f_{\tilde{N}} = \tilde{C}_{\tilde{n}_{\tilde{N}},M}\left(\rho_s,L_s;d_1,\hdots,d_{N'},I_1,\hdots,I_{N'}\right).
    \end{cases}
\end{equation}
The search domain of the design parameters is
\begin{equation}\label{eq.searchdomain}
    \begin{cases}
    D/2<d_1<d_2-D, \\
    d_1+D<d_2<d_3-D, \\
    \hspace{50pt}\vdots \\
    D+d_{N'-1}<d_{N'}<L_s/2, \\
    1\leq I^z_1\leq I^z_{\mathrm{max.}}, \\ 
    -I^z_{\mathrm{max.}}\leq I^z_2\leq I^z_{\mathrm{max.}}, \\ 
    \hspace{50pt}\vdots \\
    -I^z_{\mathrm{max.}}\leq I^z_{N'}\leq I^z_{\mathrm{max.}}, \\ 
    \end{cases}
\end{equation}
where the physical constraints on the system are that the first turn must contain a positive current, the axial turn ratios are less than the maximum axial turn ratio $I^z_{\mathrm{max.}}$, the separation of any two nested loops or arcs is less than the minimum separation $D$, and the outer loop arc is axially inside the shield $d_{N'}<L_s/2$.

We now present two examples of optimized coil designs found using a genetic algorithm. Firstly, we design an improved linear axial gradient field, $Z_2$, and compare this result to the previous anti-Helmholtz design. Then, we design a transverse bias field, $T_{1,1}$, where $M=N=1$, in which arcs are deliberately excluded from a region close to the center of the shield and compare this result to a $\cos\phi$ coil\cite{BOLINGER1989162,BIDINOSTI200531}. In both cases, we use the \textsc{MATLAB} function \texttt{gamultiobj()}, from the multi-objective genetic algorithm toolbox, which implements the NSGA-II algorithm\cite{NSGA-II} to solve for the optimal axial positions.

\subsection*{Example I: Improved linear axial gradient field}
\begin{figure}[htb]
    \centering
    \includegraphics[width=0.8\columnwidth]{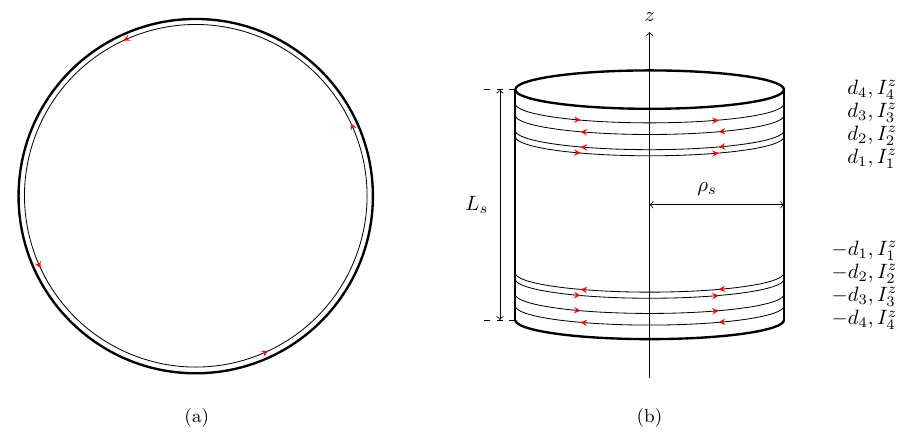}
    \caption{Schematic diagram of four anti-symmetric loop pairs of radius $\rho_s$ showing the (a) azimuthal variations and (b) axial positions $z={\pm}d_i$, where $d_i=[0.592,0.645,0.777,0.878]\rho_s$, with axial turn ratios $I^z_i=[3,-3,-2,3]$, for $i\in \mathbb{Z}:i\in[1,4]$, placed symmetrically around the origin of a closed magnetic shield of radius $\rho_s$ and length $L_s=2\rho_s$.}
    \label{fig.antihelpgapic}
\end{figure}%
\begin{figure}[htb]
    \centering
    \includegraphics[width=0.8\columnwidth]{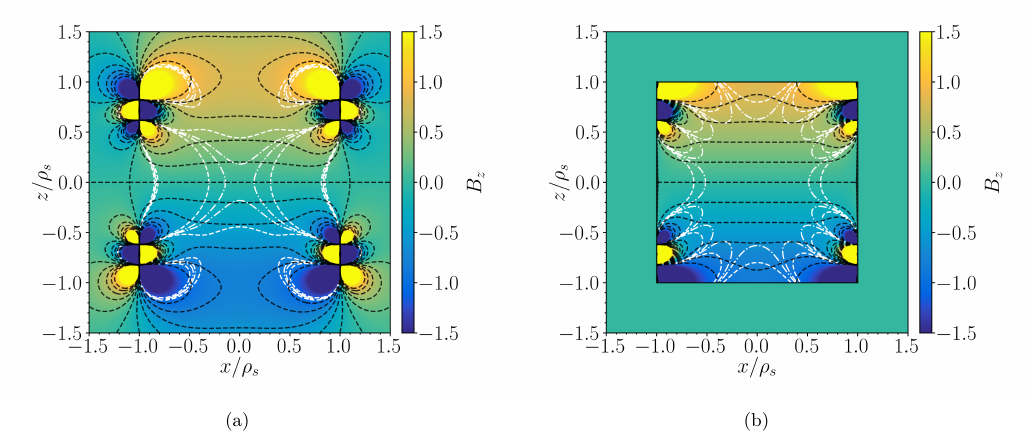}
    \caption{Color maps showing the magnitude of the normalized axial magnetic field, $B_z$, in the $xz$-plane generated the design depicted in Fig.~\ref{fig.antihelpgapic} in two situations (a) in free space and (b) placed symmetrically around the origin of a closed magnetic shield of radius $\rho_s$ and length $L_s=2\rho_s$ (solid black outline). White contours enclose the regions where the gradient of the normalized axial field with respect to $z$ deviates from unity (i.e. a perfectly uniform axial field gradient) by less than $5$\% (dashed curves) and less than $1$\% (dot-dashed curves). Black contours represent lines of constant magnetic flux (dashed curves). The resistance, gradient per unit current, and inductance of the coil both in free space and inside a unit length magnetic shield are presented in Table~\ref{table.stats}.}
    \label{fig.antihelpga}
\end{figure}%
\begin{table}[]
    \centering
    \begin{tabular}{|c| c| c c c|}
    \hline
    \multicolumn{2}{|c|}{\multirow{2}{*}{Coil Design}} & Resistance & Field / Current & Inductance \\
    \multicolumn{2}{|c|}{} & $(\Omega)$ & ($\mu$T/Am\textsuperscript{N-1}) & ($\mu$H) \\
    \hline
    \multirow{2}{3.5cm}{Anti-Helmholtz Linear Axial Gradient} & Unshielded & \multirow{2}{*}{0.269} & 3.28 & 6.19 \\ 
    & Shielded &  & 7.16 & 12.4
    \\
    \hline
    \multirow{2}{3.5cm}{Improved Linear Axial Gradient} & Unshielded & \multirow{2}{*}{2.96} & 2.74 & 131 \\
    & Shielded & & 6.92 & 184 \\
    \hline
    \multirow{2}{3.5cm}{Cosine Phi Uniform Transverse} & Unshielded & \multirow{2}{*}{2.04} & 8.32 & 278 \\
    & Shielded & & 13.4 & 595 \\
    \hline
    \multirow{2}{3.5cm}{Improved Uniform Transverse with Central Entry Region} & Unshielded & \multirow{2}{*}{3.96} & 5.26 & 496 \\
    & Shielded & & 8.58 & 716 \\
    \hline
\end{tabular}
\caption{The resistance $R$, field per unit current $C_{NM}/I$, and inductance $L$, for the example coils with wire radius $\rho_w=0.5$~mm and (standard copper) resistivity $\varrho=1.68\times10^{-8}$~${\Omega}$m, described in the text and located both in free space and inside a magnetic shield of unit diameter and length, $\rho_s=0.5$~m and $L_s=1$~m, respectively. The anti-Helmholtz and improved linear axial gradient coils are shown in Fig.~\ref{fig.optahelmpic} and Fig.~\ref{fig.antihelpgapic}, respectively, and generate an $N=2$ zonal harmonic field, $Z_2$. The cosine phi ($\cos \phi$) and improved uniform transverse coils are shown in Fig.~\ref{fig.transpic} and Fig.~\ref{fig.transgapic}, respectively, and generate an $N=1$, $M=1$ tesseral harmonic field, $T_{1,1}$. The inductance is calculated numerically using \textsc{COMSOL Multiphysics}\textsuperscript{\textregistered} Version 5.5.}
\label{table.stats}
\end{table}
To find an improved linear axial gradient field we choose to search for solutions using a four-pair anti-symmetric loop setup within a high-permeability magnetic shield of aspect ratio $L_s/(2\rho_s)=1$. The axial magnetic field is given by 
\begin{align}\label{eq.bzgradgen}
    B_{z}\left(\rho,\phi,z\right)=\frac{2\mu_0}{L_s}\sum_{i=1}^{4}I^z_ib^-_0(\rho,z;d_i),
\end{align}
and we choose to minimize the first three leading-order error terms
\begin{equation}\label{eq.ex1obj}
    \begin{cases}
     \min f_1 = \tilde{C}_{4,0}\left(\rho_s,L_s;d_1,\hdots,d_4,I^z_1,\hdots,I^z_4\right), \\
     \min f_2 = \tilde{C}_{6,0}\left(\rho_s,L_s;d_1,\hdots,d_4,I^z_1,\hdots,I^z_4\right), \\
     \min f_3 = \tilde{C}_{8,0}\left(\rho_s,L_s;d_1,\hdots,d_4,I^z_1,\hdots,I^z_4\right),
    \end{cases}
\end{equation}
which, from \eqref{eq.iBzg}, are given by
\begin{align}
    \tilde{C}_{2n,0}=\sum_{i=1}^4 \frac{I_i^z}{L_s^{2n-1}}\sum_{p=1}^{\infty} (2p-1)^{2n-1}\frac{\sin\left(\frac{\pi d_i(2p-1)}{L_s}\right)}{I_{0}\left(\frac{\pi(2p-1)\rho_s}{L_s}\right)}.
\end{align}
We constrain the separation of the wires such that $D=0.01\rho_s$, limit the maximum turn ratio to $I^z_{\mathrm{max.}}=9$, assume a wire radius $\rho_w=0.001\rho_s$, and search for optimal values of $[d_1,\hdots,d_4]$ and $[I_1^z,\hdots,I_4^z]$. The genetic algorithm outputs numerous solutions where the first three undesired contributions are minimized, meaning that many solutions exist where no undesired harmonic can be further minimized without increasing the magnitude of another undesired harmonic. To filter these solutions, we first discard solutions where all three harmonics are insufficiently nulled. Then, we rank the remaining solutions according to their stability by adjusting each wire placement in turn by $\pm\rho_w$ and analyzing the magnitude of the leading-order error terms. In appendix B, we describe the implementation of \texttt{gamultiobj()} to the minimization of the effective harmonic magnitudes and benchmark its performance. Averaged over ten runs, the optimization takes $5.86$ s and requires $127500$ evaluations of each objective function, \eqref{eq.ex1obj}.

The optimized coil configuration is shown in Fig.~\ref{fig.antihelpgapic}. The color maps in Fig.~\ref{fig.antihelpga} show the magnitude of the axial magnetic field component, $B_z$, generated (a) in free space and (b) inside the high-permeability magnetic shield. Due to the improved linearity of the magnetic field profile that results from the additional coil pairs, the volume of the region within which the field achieved is within $1$\% of the desired field (i.e. within the dot-dashed curves) is seven times larger than that produced by standard anti-Helmholtz coils inside the same magnetic shield (see Fig.~\ref{fig.inshieldhomo}). This demonstrates the effectiveness of our design methodology and the applicability of the genetic algorithm optimization to this problem. The resistance, magnetic field gradient per unit current, and inductance for various coil configurations in both free space and inside a magnetic shield of unit length are summarized in Table.~\ref{table.stats}. The resistance and inductance of the optimized coil is, however, an order of magnitude larger than for the standard anti-Helmholtz configuration. Additional constraints could be added to the optimization to minimize the resistance, inductance, or to maximize the field per current, i.e. by reducing $I^z_{\mathrm{max.}}$, imposing that all currents must flow with the same parity, or adding a constraint to maximize the magnitude of the desired harmonic. However, these would come at a cost of field fidelity. We also present a comparison of this design to a coarse discretization of an optimized continuum current distribution on the surface of a cylinder\cite{PhysRevApplied.14.054004} in Appendix C.

\subsection*{Example II: Improved uniform transverse field with a central entry region}
\begin{figure}[b!]
    \centering
    \includegraphics[width=0.8\columnwidth]{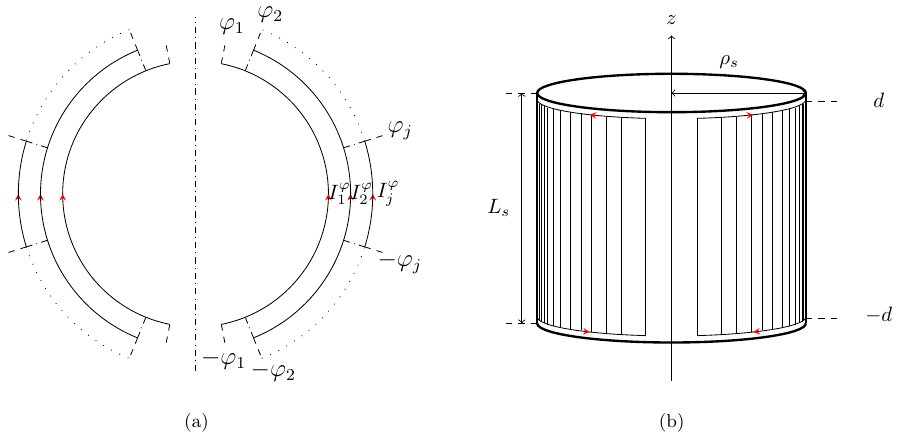}
    \caption{Schematic diagram of a single axially anti-symmetric arc pair of radius $\rho_s$ with $M'=12$ azimuthal variations, generating a saddle-like $\cos \phi$ coil from references~\citenum{BOLINGER1989162}~and~\citenum{BIDINOSTI200531}, showing the (a) azimuthal variations of periodicity, $\pi$, for the twelve separate angular lengths $\varphi_j=\left[\arccos\left(1-\left(j-\frac{1}{2}\right)/M'\right)\right]$, for $j\in \mathbb{Z}:j\in[1,M']$, each with an azimuthal turn ratio of unity, and (b) axial position $z={\pm}d$, where $d=(L_s/2-\rho_w)\rho_s$, placed symmetrically around the origin of a closed magnetic shield of radius $\rho_s$ and length $L_s=2\rho_s$.}
    \label{fig.transpic}
\end{figure}%
\begin{figure}
    \centering
    \includegraphics[width=0.8\columnwidth]{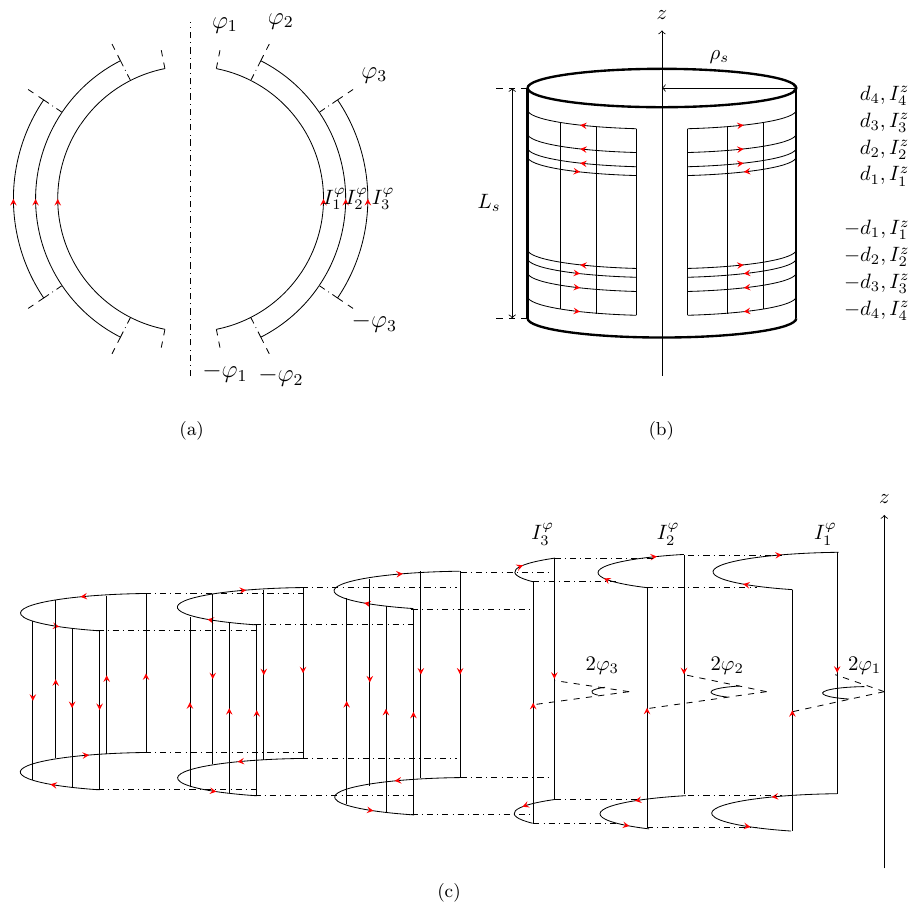}
\caption{Schematic diagram of four axial anti-symmetric arc pairs of radius $\rho_s$ with three azimuthal variations for each pair, showing the (a) azimuthal variations of periodicity, $\pi$, for the three separate angular lengths $\varphi_j=[1.367,1.101,0.592]$ with azimuthal turn ratios $I^\varphi_j=[1,1,1]$, for $j\in \mathbb{Z}:j\in[1,3]$, and (b) axial positions $z={\pm}d_i$, where $d_i=[0.600,0.651,0.781,0.938]\rho_s$, with axial turn ratios $I^z_i=[4,-2,-2,-1]$, for $i\in \mathbb{Z}:i\in[1,4]$, placed symmetrically around the origin of a closed magnetic shield of radius $\rho_s$ and length $L_s=2\rho_s$. (c) Shows an expanded schematic diagram of one azimuthal section of the coil depicted in (a)-(b) for clarity.}
\label{fig.transgapic}
\end{figure}%
\begin{figure}
    \centering
    \includegraphics[width=0.8\columnwidth]{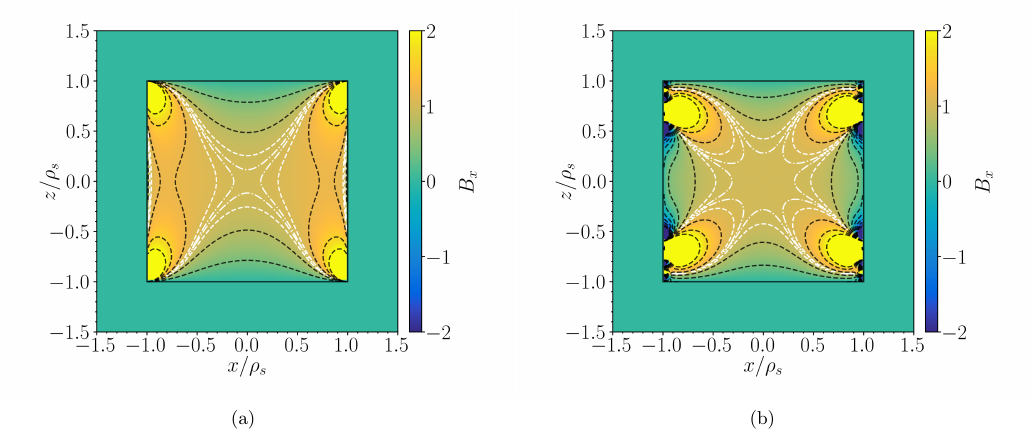}
    \caption{Color maps showing the magnitude of the normalized transverse magnetic field, $B_x$, in the $xz$-plane generated by (a) the $\cos~\phi$ uniform $B_x$ field-generating design depicted in Fig.~\ref{fig.transpic} and (b) the optimized uniform $B_x$ field-generating design depicted in Fig.~\ref{fig.transgapic}, placed symmetrically around the origin of a closed magnetic shield of radius $\rho_s$ and length $L_s=2\rho_s$ (solid black outline). White contours enclose the regions where the normalized transverse field deviates from unity (i.e. a perfectly uniform transverse field) by less than $5$\% (dashed curves) and less than $1$\% (dot-dashed curves). Black contours represent lines of constant magnetic flux (dashed curves). The resistance, field per unit current, and inductance of the coil both in free space and inside a unit length magnetic shield are presented in Table~\ref{table.stats}.}
    \label{fig.trans_new}
\end{figure}%
Now, we design a uniform transverse field, $B_x$, which can be represented by a single spherical harmonic field of order $N=1$ and degree $M=1$. Consequently, the symmetries within the desired harmonic field correspond to the anti-symmetric tesseral coil basis, with azimuthal periodicity $\pi$, as shown in Fig.~\ref{fig.symanti}d and Fig.~\ref{fig.arcs}b. As mentioned above, the harmonic $T_{1,1}$ is not present within the axial field. However, we can still search for optimized transverse coils using the axial magnetic field. Here, we use a setup comprising four pairs of coils with three overlapping arcs of different angular lengths, which generate an axial magnetic field
\begin{align}\label{eq.bxbiasgen}
    B_{z}\left(\rho,\phi,z\right)=\frac{4\mu_0}{L_s}\sum_{m=1}^{\infty}\sum_{i=1}^{4}\sum_{j=1}^{3}I^z_iI^\varphi_jb^-_m(\rho,z;d_i)\Phi^m_j\cos(m\phi),
\end{align}
where 
\begin{equation}
    \Phi^m(\varphi_j)=\frac{\sin(m\varphi_j)}{\pi m}(1-(-1)^m).
\end{equation}
Using three angular lengths, we can remove the first three sets of harmonics of degrees $m=\left(3,5,7\right)$ by solving the set of simultaneous equations
\begin{equation}\label{eq.phimintrans}
    \min_{\varphi_j}\left(\sum_{j=1}^{M'}I^\varphi_j\sin((2\nu+1)\varphi_j)\right), \qquad \nu\in \mathbb{Z}:\nu\in[1,3].
\end{equation}
For simplicity and ease of manufacturing, we choose $I_j^\varphi=[1,1,1]$, and find optimized angular lengths of $\varphi_j=[1.367,1.101,0.592]$ to remove the leading-order azimuthal variations of degrees $m=\left(3,5,7\right)$. The angular lengths are calculated in $0.70$~ms using the \texttt{FindRoot[]} function in \textsc{Mathematica}. 

Having removed the first three leading-order azimuthal variations in the desired field, the first three leading-order error terms in the total field are given by
\begin{align}\label{eq.CnmBy}
    \tilde{C}_{2n+1,1}=\sum_{i=1}^N \frac{I^z_i}{L_s^{2n}} \sum_{p=1}^{\infty} (2p-1)^{2n}\frac{\sin\left(\frac{\pi d_i(2p-1)}{L_s}\right)}{I_{1}\left(\frac{\pi(2p-1)\rho_s}{L_s}\right)}, \qquad n\in \mathbb{Z}:n\in[1,3],
\end{align}
where the objective functions are written as
\begin{equation}\label{eq.ex2obj}
    \begin{cases}
     \min f_1 = \tilde{C}_{3,1}\left(\rho_s,L_s;d_1,\hdots,d_4,I_1,\hdots,I_4\right), \\
     \min f_2 = \tilde{C}_{5,1}\left(\rho_s,L_s;d_1,\hdots,d_4,I_1,\hdots,I_4\right), \\
     \min f_3 = \tilde{C}_{7,1}\left(\rho_s,L_s;d_1,\hdots,d_4,I_1,\hdots,I_4\right).
    \end{cases}
\end{equation}
Again, we impose the constraints, $D=0.01\rho_s$, $I^z_{\mathrm{max.}}=9$, and $\rho_w=0.001\rho_s$, and search for optimal values of $[d_1,\hdots,d_4]$ and $[I_1^z,\hdots,I_4^z]$. Additionally, we shall impose a constraint that $d_1=3L_s/10$ so that optical access is maintained inside large windows near the axial origin, e.g. for laser/electronic access. Following the method described in example I, the most stable Pareto-optimal solution is selected. This effectively eliminates the first two leading-order error terms and greatly reduces the third. The optimization takes $23.6$ s and requires $526000$ evaluations of each objective function, \eqref{eq.ex2obj}. Here, we note that, to minimize the set of spatial variations as efficiently as possible, the number of azimuthal degrees nulled is matched to the leading-order axial variation which is not nulled, i.e. where the leading-orders $n=\left(3,5,7\right)$ are minimized, nulling the degrees $m=\left(3,5,7\right)$ is appropriate.

We can compare the performance of the optimized coil to a standard discrete saddle-shaped $\cos\phi$~coil\cite{BOLINGER1989162} within the same magnetic shield. A discrete $\cos\phi$~coil is a specific case of the $M=1$ anti-symmetric coil basis. This coil is constructed of $M'$ pairs of nested single saddles\cite{BIDINOSTI200531} of angular lengths $\varphi_j=\left[\arccos\left(1-\left(j-\frac{1}{2}\right)/M'\right)\right]$, for $j\in \mathbb{Z}:j\in[1,M']$, each with an azimuthal turn ratio of unity. The set of axial wires which make up the saddles emulate the axial current density $J_z\left(\phi',z'\right)=\cos\phi'$. When the respective angular lengths are substituted into \eqref{eq.phimin}, in the case where $M'{\to}\infty$ all undesired degrees are minimized except $M=1$. The axial separation, $d=L_s-\rho_w$, of the saddles is extended as far as possible along the whole length of the shield to minimize the leading-order error harmonic, $\tilde{N}=3$. This can be demonstrated by substituting $n=1$ and $d=L_s-\rho_w$ into \eqref{eq.CnmBy}, and noting the $\tilde{N}=3$ error harmonic is nulled to zero for $\rho_w{\to}0$. Here, to make a fair comparison between the $\cos\phi$~coil and the optimized design, we set the number of pairs of nested saddles, $M'=12$, equal to the number of sets of saddles in the optimized design.

The wire configurations of the $\cos\phi$~coil and the optimized coil are presented in Figs.~\ref{fig.transpic}~and~\ref{fig.transgapic} and their coil properties are summarized in Table~\ref{table.stats}. The transverse field variations in the $xz-$plane inside the magnetic shield generated by the $\cos\phi$~coil and optimized coil are shown in Fig.~\ref{fig.trans_new}. The optimized coil contains windows for optical access along the axial center of the shield. These windows extend over $60$\% of the shield's length and are more than twice as great in azimuthal extent than the equivalent spaces in the $\cos\phi$~coil. The optimized transverse coil generates a field that is homogeneous to within $1\%$ variation throughout a volume that is approximately three times greater than that generated by the $\cos\phi$~coil. However, the resistance and inductance for the optimized transverse field coils are, respectively, $1.9$ and $1.2$ times larger than the $\cos\phi$~coil. The field per unit current is also a factor of $1.6$ lower in the optimized system compared with the $\cos\phi$~coil. As with the previous example, additional constraints could be added to the optimization to improve the desired coil properties, at the likely cost of some field fidelity.

\section{Conclusion}
In summary, we have introduced a coil design method based around simple discrete current-carrying loops and arcs whose geometry can be optimized to generate any physically-attainable magnetic field within a high-permeability cylindrical magnetic shield to a high fidelity. To do this, we determined field expansions that enable elimination of deviations from the desired field to a specified expansion order when the coil is on the magnetic shield's surface. We then presented a discrete coil basis composed of unit-coil building blocks and decomposed the magnetic field into spherical harmonic terms in free space. Next, for specific designs, we related the coil parameters, namely the wire spacing, angular arc lengths, and the currents through pairs of loops and arcs, to a set of harmonic fields chosen to reflect the form of the desired field profiles. We then used our model to determine the variation in the optimal separation of an anti-Helmholtz pair in magnetic shields of different aspect ratios. Taking this optimization one step further, we formulated simultaneous equations to remove multiple harmonic fields using multiple current loops and arcs and used a genetic algorithm to find optimized turn ratios and wire separations. We used this optimization procedure to design high-fidelity transverse bias and linear-gradient fields. In particular, we found that this optimization process increased the volume within which variations of the field gradient are less than $1\%$ by a factor of seven compared to the standard anti-Helmholtz arrangement in the same shield.

Both the harmonic magnitudes and the Fourier series representation of the field generated by each building block can be calculated rapidly, enabling multiple functional evaluations during the design process. Moreover, the discrete coil basis is additive, meaning that building block units can be added to, or removed from, a coil depending on the required performance. This methodology will facilitate new miniaturized technologies that require custom magnetic fields within a magnetically shielded environment. The performance of existing magnetic field-generating systems can be improved by retrofitting discrete coil systems that are optimized by our methodology. Additional objective functions could be added to the method to maximize the desired harmonic, minimize inductance, or test the representation of the desired harmonic under shifts in wire placement. Further research could investigate the use of discrete planar coils on the surface of the end-plates to enable more power-efficient and high-fidelity designs. As well as this, one could consider analytical solutions for the electromagnetic coupling of either the spherical coil basis or projected spherical coil basis to magnetic shields of various topologies.

\section*{Acknowledgments} 
We acknowledge support from the UK Quantum Technology Hub in Sensors and Timing, funded by the UK Engineering and Physical Sciences Research Council (EP/M013294/1) and from Innovate UK Project 44430 MAG-V: Enabling Volume Quantum Magnetometer Applications through Component Optimisation \& System Miniaturisation. 

\section*{Author Declarations} 
\subsection*{Conflict of interest}
The authors M.P., P.J.H., T.M.F., M.J.B., and R.B. declare that they have a patent pending to the UK Government Intellectual Property Office (Application No.1913549.0) regarding the magnetic field optimization techniques described in this work. N.H, M.J.B, and R.B also declare financial interests in a University of Nottingham spin-out company, Cerca. N.L.H is a paid employee of Wolfram Research, who make \textsc{Mathematica}. The authors have no other conflicts to disclose.

\section*{Data Availability Statement} 
The data that support the findings of this study are available from the corresponding author upon reasonable request. Verification using \textsc{MATLAB}, \textsc{Mathematica}, or \textsc{COMSOL Multiphysics}\textsuperscript{\textregistered} requires a valid license. All calculations were performed using the CPU of a MacBookPro16,1 containing a 6-Core Intel Core i7 2.6 GHz processor with 16 GB of DDR4 RAM.

\section*{References} 

\bibliographystyle{apsrev4-2}

\newpage

\section*{Appendix A: Axial Differentiation of Spherical Harmonics}
\renewcommand{\theequation}{A.\arabic{equation}}
\renewcommand{\thesection}{A.\arabic{section}}  
\setcounter{equation}{0}   

Let us consider the harmonic 
 \begin{equation}\label{eq.phis}
     R_{n,m}(r,\theta,\phi)=r^nP_{n,|m|}\left(\cos\theta\right)
     \begin{pmatrix}
    \cos\left(|m|\phi\right)\\
    \sin\left(|m|\phi\right)
    \end{pmatrix}.
    \qquad
    \begin{matrix}
    m\geq0\\
    m<0
\end{matrix}
 \end{equation}
 The differential of an arbitrary curvilinear coordinate system with respect to another may be expressed as
 \begin{equation}\label{eq.curv}
     \frac{\partial  R_{n,m}(r,\theta,\phi)}{\partial \chi_i}=\sum_j\frac{\partial \xi_j}{\partial \chi_i}\frac{\partial }{\partial \xi_j} R_{n,m}(r,\theta,\phi).
 \end{equation}
Using this, the differential in cylindrical coordinates may be determined. Spherical polar coordinates can be written as 
 \begin{align}\label{eq.coord}
 r=\sqrt{\rho^2+z^2}, \qquad
 \theta=\cos^{-1}\left(\frac{z}{\sqrt{\rho^2+z^2}}\right), \qquad
 \phi=\phi.
 \end{align}
 As a result, the axial derivative is given by
 \begin{equation}\label{eq.partial}
     \frac{\partial R_{n,m}(r,\theta,\phi)}{\partial z}=\left(\frac{\partial r}{\partial z}\frac{\partial }{\partial r}+\frac{\partial \theta}{\partial z}\frac{\partial }{\partial \theta}\right)R_{n,m}(r,\theta,\phi),
 \end{equation}
Using \eqref{eq.curv} and \eqref{eq.coord}, the differential with respect to $z$ is given simply by
\begin{equation}
    \frac{\partial R_{n,m}(r,\theta,\phi)}{\partial z}=\left(\cos\theta\frac{\partial}{\partial r}-\frac{\sin\theta}{r}\frac{\partial}{\partial \theta}\right)R_{n,m}(r,\theta,\phi),
\end{equation}
which, using \eqref{eq.phis}, becomes
\begin{equation}\label{eq.diffphi}
    \frac{\partial R_{n,m}(r,\theta,\phi)}{\partial z}=r^{n-1}\left(n\cos\theta P_{n,|m|}\left(\cos\theta\right)-\sin\theta\frac{\partial P_{n,|m|}\left(\cos\theta\right)}{\partial \theta}\right)
    \begin{pmatrix}
    \cos\left(|m|\phi\right)\\
    \sin\left(|m|\phi\right)
    \end{pmatrix}.
    \qquad
    \begin{matrix}
    m\geq0\\
    m<0
\end{matrix}
\end{equation}
Directly substituting the relation from reference~\citenum{mathsbook}
\begin{equation}
    \frac{\partial P_{n,|m|}\left(\cos\theta\right)}{\partial \theta}=n\cot\theta P_{n,|m|}\left(\cos\theta\right)-\frac{n+|m|}{\sin\theta}P_{n-1,|m|}\left(\cos\theta\right)
\end{equation}
into \eqref{eq.diffphi} yields the final expression
\begin{equation}
    \frac{\partial R_{n,m}(r,\theta,\phi)}{\partial z}=(n+|m|)r^{n-1} P_{n-1,|m|}\left(\cos\theta\right)     \begin{pmatrix}
    \cos\left(|m|\phi\right)\\
    \sin\left(|m|\phi\right)
    \end{pmatrix}.
    \qquad
    \begin{matrix}
    m\geq0\\
    m<0
\end{matrix}
\end{equation}

\section*{Appendix B: Implementation and Benchmarking of the Genetic Algorithm}
\renewcommand{\theequation}{B.\arabic{equation}}
\renewcommand{\thesection}{B.\arabic{section}}  
\setcounter{equation}{0}   
\begin{figure}[htb]
    \centering
    \includegraphics[width=0.8\columnwidth]{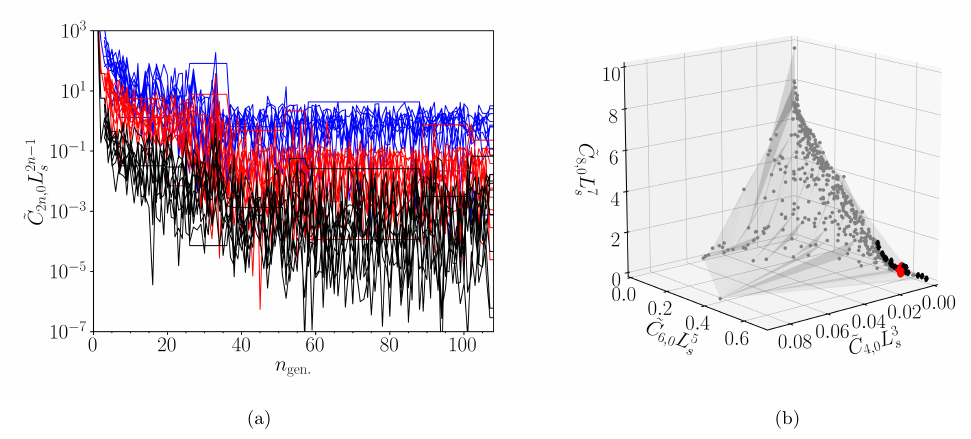}
    \caption{(a-b) Implementation of the genetic algorithm to design the improved linear axial field gradient displayed in Fig.~\ref{fig.antihelpgapic} inside a magnetic shield of radius $\rho_s$ and length $L_s=2\rho_s$. (a) The effective magnitude of the first three scaled leading-order error harmonics, $\tilde{C}_{2n,0}L_s^{2n-1}$, where $n=2$ is the cubic gradient (black), $n=3$ is the quintic gradient (red), and $n=4$ is the septic gradient (blue), of the ten randomly-selected members of the population with $N_\mathrm{pop.}=1000$ members as the number of generations, $n_{\mathrm{gen.}}\in \mathbb{Z}:i\in[1,108]$, progresses. Convergence is achieved after $108$ generations. (b) Pareto front (grey shaded and scatter) on which the first three leading-order effective harmonic magnitudes are minimized. Filtered solutions and the most stable solution to minimize the cubic gradient are highlighted (black and red, respectively).}
    \label{fig.geneticalgorithm}
\end{figure}%
Here, we provide information about the implementation of the genetic algorithm to solve for the optimal coil geometries and provide its performance specification for the examples presented in the main text. We use the NSGA-II algorithm\cite{NSGA-II} as implemented using the \texttt{gamultiobj()} function in the \textsc{MATLAB} Global Optimization Toolbox. This algorithm is simple to implement and, importantly, is elitist and controlled, meaning that it prioritizes members of the population that are functionally closer to the objective and improve the diversity of the total population, respectively. The algorithm is used simultaneously to minimize multiple axial variations, \eqref{eq.gaobj}, by determining optimal axial positions and turn ratios subject to constraints on the search domain, \eqref{eq.searchdomain}. We modify the base mutation, crossover, and creation functions in \texttt{gamultiobj()} so that integer turn ratios and continuous axial separations can be optimized simultaneously.

In the examples presented in the main text, we wish to generate a linear axial field gradient and a uniform transverse field by minimizing \eqref{eq.ex1obj} and \eqref{eq.ex2obj}, respectively. The crossover rate and Pareto fraction are set to standard values of $0.9$ and $0.5$, respectively\cite{GACross}. As the higher-order effective harmonic magnitudes are very sensitive to small changes in the geometric input variables, we initialize the variables randomly and use shrink mutation with default parameters\cite{ShrinkMutation}. In addition, we use a large population size, $N_\mathrm{pop.}=1000$, to enhance the exploration of the optimization landscape\cite{GAbook}. We use a standard number of maximum generations, $N_\mathrm{gen.}=10N_\mathrm{pop.}$, and stop the algorithm if the spread, i.e. the movement of the solutions on the Pareto front, is smaller than a standard\cite{NSGA-IIBench} NGSA-II function tolerance, $1\times10^{-4}$, over a standard number of stall generations, $100$. To encode the search domain, \eqref{eq.searchdomain}, the maximal and minimal bounds of each input variable are imposed as lower and upper bounds and, additionally, the minimum separation between adjacent loops is imposed as a linear inequality constraint.

In Fig.~\ref{fig.geneticalgorithm}a, we plot the effective magnitudes of the scaled first, second, and third leading-order error harmonics of ten randomly-selected members of the population at each generation in the design of the improved linear axial field gradient coil. The axial variations are scaled so that they are dimensionless quantities applicable to design in any shield with aspect ratio $L_s/(2\rho_s)=1$ via appropriate adjustment of the applied current. An example Pareto front on which these axial variations are minimized is presented in Fig.~\ref{fig.geneticalgorithm}b. As described in the main text, we filter the solutions on the Pareto front according to how effectively the harmonics are minimized and then rank solutions according to their stability. In this case, we choose this filtering to be $\tilde{C}_{4,0}L_s^3<10^{-4}$, $\tilde{C}_{6,0}L_s^5<1$, and $\tilde{C}_{8,0}L_s^7<1$ (black in Fig.~\ref{fig.geneticalgorithm}b). We rank the stability of solutions by adjusting each wire placement in turn by $\pm\rho_w$ and selecting the solution which minimizes the sum of the proportionate increases in each of the leading-order error harmonics. It should also be noted that, when we run the algorithm numerous times, there exist other solution modes which may manifest themselves after the ranking since they also null the sum of harmonics near-totally and are stable. In this case, we choose the solution with the lowest sum of absolute turn ratio magnitudes. Alternatively, nulling of the fourth leading-order error harmonic or maximization of the desired harmonic could also be used. The solution presented in the main text (red in Fig.~\ref{fig.geneticalgorithm}b) is then rounded to three decimal places since positioning below $1$~mm precision is impractical. Averaged over ten runs, the optimization takes $5.86$ s and requires $127500$ evaluations of each objective function, \eqref{eq.ex1obj}. Averaged over these ten runs, for the optimal solution mode, the standard errors in the axial positions, $\alpha\left({d_i}\right)=[0.0002,0.0007,0.0009,0.003]\rho_s$, are below the precision to which we quote the axial positions in the main text.

Now, let us compare the performance of the algorithm to an exhaustive search. We set the range of axial positions of the loops coarsely to $d_i=0.05j\rho_s$ for $j\in \mathbb{Z}:i\in[1,19]$, meaning that there are $3876$ unique combinations of the four axial positions after the conditions on the search domain, \eqref{eq.searchdomain}, are applied. Using $I_{\mathrm{max.}}^z=9$, there are $9$ allowed integer $I_z^{1}$ and $19$ allowed $I_z^{j}$ for $j\in[2,4]$, giving $61731$ unique combinations of currents. Combining these parameter conditions requires us to evaluate the objective function $239269356$ times, over $1800$ times as many iterations as was used in the genetic algorithm. This takes $148$ minutes to evaluate, and no solution is found which minimizes the objective functions to match the filtering conditions ($\tilde{C}_{4,0}L_s^3<10^{-4}$, $\tilde{C}_{6,0}L_s^5<1$, and $\tilde{C}_{8,0}L_s^7<1$). Clearly, the algorithm is more robust and computationally efficient than an exhaustive search. Future investigations could compare the computational efficiency and robustness of the genetic algorithm to other multi-objective optimization routines, such as particle swarm optimization and simulated annealing.

The design of the uniform transverse field follows a similar implementation to the improved linear axial field gradient. Compared to the previous example, the objective functions have increased spatial variability. This means that the algorithm requires more function evaluations to reach its stopping condition. The optimization takes $23.6$ seconds and requires $526000$ evaluations of each objective function.

\section*{Appendix C: Comparison between Building Block and Continuum Linear Axial field Gradient Designs}
\renewcommand{\theequation}{C.\arabic{equation}}
\renewcommand{\thesection}{C.\arabic{section}} 
\begin{figure}[htb]
    \centering
    \includegraphics[width=0.8\columnwidth]{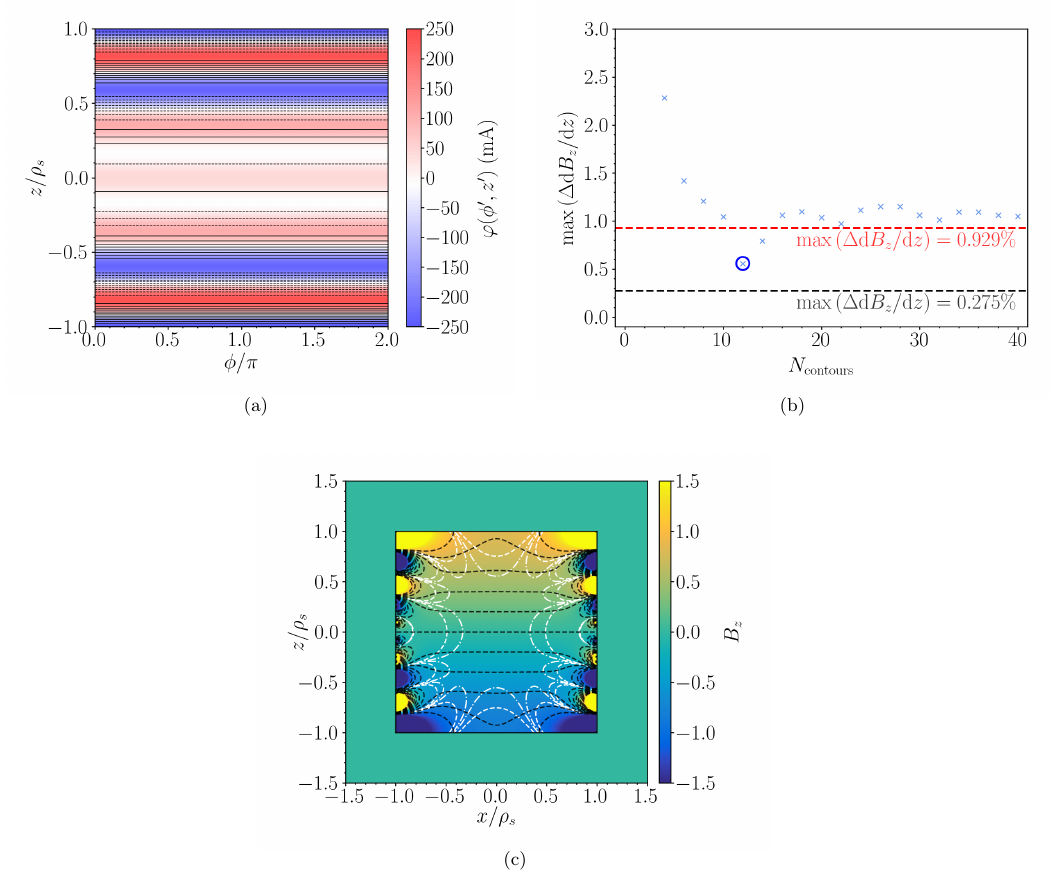}
    \caption{Schematic diagram (a) and performance (b-c) of a linear axial field gradient, $\mathrm{d}B_z/\mathrm{d}z$, generating coil system inside a closed magnetic shield of radius $\rho_s$ and length $L_s=2\rho_s$, where a discretized representation of a continuum of current is housed open cylinder placed symmetrically on the inner surface of the shield and of the same dimension as the shield. (a) Streamfunction of the continuum current, where positive to negative values are represented from red to white to blue, respectively. Black curves show a discretized representation of the continuum current with $N_\mathrm{contours}=12$ contour levels. Opposite current flow directions are represented with solid and dashed line-styles, respectively. This coil was designed using the method of Ref.~[28], with optimization parameters $N=100$, $M=0$, $\beta=5.95\times10^{-15}$~T\textsuperscript{2}/W, $t=0.5$~mm, and $\rho=1.68\times10^{-8}$~$\Omega$m. (b) Maximum deviation in the normalized axial field gradient, $\mathrm{max}\left(\Delta\mathrm{d}B_z/\mathrm{d}z\right)$, evaluated over the the central half of the $z$-axis, $z=[-L_s/4,L_s/4]$, calculated for discrete representations of the continuum of current as the number of contour levels is increased (blue scatter). The maximum deviations calculated for the continuum of current in (a) (red dotted) and design in Fig.~\ref{fig.antihelpgapic} (black dotted) are also plotted in the same context. The discrete representation which minimizes the error is found at $N_\mathrm{contours}=12$ (dark blue circle). (c) $B_z$ in the $xz$-plane generated the discrete representation with $N_\mathrm{contours}=12$ inside the magnetic shield. White contours enclose the regions where the gradient of the normalized axial field with respect to $z$ deviates from unity (i.e. a perfectly uniform axial field gradient) by less than $5$\% (dashed curves) and less than $1$\% (dot-dashed curves). Black contours represent lines of constant magnetic flux (dashed curves).}
    \label{fig.antihelcontinuum}
\end{figure}%

Here, we design a magnetic field coil using a discretized continuum current density on the open cylindrical inner surface of a closed cylindrical magnetic shield. First, the azimuthal current density is posed in a Fourier basis, which is substituted into~\eqref{eq.brdisbig}-\eqref{eq.bzdisbig} and solved analytically. Details about this may be found in Ref~[28]. As current is conserved on the coil surface, the azimuthal current density can then be related to the gradient of a streamfunction\cite{jackson}. To manufacture any design from the continuum solution, a discretized representation of the continuum current is generated by contouring the streamfunction at an even number of evenly separated levels, $N_\mathrm{contours}$. As discussed in the main text, when the continuum is coarsely discretized, i.e. $N_\mathrm{contours}$ is low, the magnetic field generated by the discrete wire pattern may be substantially different from that generated by the continuum. Thus, in cases where the $N_\mathrm{contours}$ must be low, e.g. a miniaturized device, a building block design may be preferable. This is because the coarsely discretized continuum coil will require a thorough discretization analysis. Consequently, this analysis can only be determined \emph{a posteriori} unless the coil topology is highly idealized. The discretized patterns can have complex topologies and may require FEM software to evaluate\cite{Hobson2021BespokeMF}, which may be computationally intensive. However, because zonal harmonics have total azimuthal symmetry, the contours may be represented as loop pairs. This means that the magnetic field generated by discretized zonal patterns can now be evaluated analytically using~\eqref{eq.bztotal}.

In Fig.~\ref{fig.antihelcontinuum}a, we present a continuum coil designed to generate a linear axial gradient field, $\mathrm{d}B_z/\mathrm{d}z$, inside a magnetic shield of aspect ratio $L_s=2\rho_s$. The regularization in this design is deliberately low such that the field fidelity is as high as possible. This, however, comes at the cost of a highly oscillatory streamfunction (see Ref~[28]). In Fig.~\ref{fig.antihelcontinuum}b, the maximum axial field gradient error, $\mathrm{max}\left(\Delta\mathrm{d}B_z/\mathrm{d}z\right)$, over the central $50$\% of the $z$-axis from the center of the magnetic shield is calculated as the number of contour levels is increased (blue scatter). As the number of contours increases, the maximum field gradient error tends to that calculated from the continuum current (red) with some small offset due to the difficulty in representing very high spatial frequency features in the design. Notably, however, for $N_\mathrm{contours}=12$ (highlighted with dark blue circle), the calculated maximum field gradient error is significantly reduced compared to the other cases. The axial field variations in the $xz-$plane inside the magnetic shield generated by this discretized continuum coil are shown in Fig.~\ref{fig.antihelcontinuum}c.

We now compare the optimally discretized continuum design with $N_\mathrm{contours}=12$ to the building block improved linear axial gradient field design in the main text (Fig.~\ref{fig.antihelpgapic}). The volume of the central region on the $xz-$plane within which the field achieved is within $1$\% of the desired field is a factor of $0.76$ smaller for the discretized continuum design compared to the building block design (Fig.~\ref{fig.antihelpga}b). This means that the optimally discretized continuum coil represents the desired field profile to a slightly reduced fidelity. Generally, building block designs will have fewer unique wire placements than optimally discretized continuum coils. In this case, the optimally discretized continuum coil contains wire pairs at $35$ unique positions, whereas the building block coil has $4$ unique positions; this increases the inductance by a factor of $6.4$ (Table~\ref{table.stats}) and may also block optical access. On the other hand, the process of strictly minimizing undesired harmonics while not controlling the objective harmonic may reduce the relative field per unit current of the building block coil. In this case, the optimally discretized continuum coil has a field per unit current a factor of $2.1$ times greater at the center of the shield (Table~\ref{table.stats}).

\end{document}